\documentclass[aps,superscriptaddress,twocolumn,longbibliography]{revtex4-1}

\usepackage[pdftex]{graphicx}
\usepackage{dcolumn}
\usepackage{bm}
\usepackage{color}
\usepackage{amsmath}
\usepackage{nicefrac}
\usepackage{amssymb} 

\newcommand{\ff}[1]{{\boldsymbol #1}}
\newcommand{\ca}[1]{{\cal #1}}
\newcommand{\bi}{\begin{itemize}}
\newcommand{\ei}{\end{itemize}}
\newcommand{\be}{\begin{equation}}
\newcommand{\ee}{\end{equation}}
\newcommand{\ba}{\begin{eqnarray}}
\newcommand{\ea}{\end{eqnarray}}

\usepackage[pdftex,colorlinks=true,linkcolor=blue,citecolor=blue,filecolor=blue]{hyperref}

\begin{document} 
  
\title{Pre-relaxation in quantum, classical, and quantum-classical two-impurity models}

\author{Michael Elbracht} 

\affiliation{I.\ Institute of Theoretical Physics, Department of Physics, University of Hamburg, Notkestra\ss{}e 9, 22607 Hamburg, Germany}

\author{Michael Potthoff}

\affiliation{I.\ Institute of Theoretical Physics, Department of Physics, University of Hamburg,  Notkestra\ss{}e 9, 22607 Hamburg, Germany}

\affiliation{The Hamburg Centre for Ultrafast Imaging, Luruper Chaussee 149, 22761 Hamburg, Germany}

\begin{abstract}
We numerically study the relaxation dynamics of impurity-host systems, focusing on the presence of long-lived metastable states in the non-equilibrium dynamics after an initial excitation of the impurities.
In generic systems, an excited impurity coupled to a large bath at zero temperature is expected to relax and approach its ground state over time. 
However, certain exceptional cases exhibit metastability, where the system remains in an excited state on timescales largely exceeding the typical relaxation time. 
We study this phenomenon for three prototypical impurity models: a tight-binding quantum model of independent spinless fermions on a lattice with two stub impurities, a classical-spin Heisenberg model with two weakly coupled classical impurity spins, and a tight-binding quantum model of independent electrons with two classical impurity spins. 
Through numerical integration of the fundamental equations of motion, we find that all three models exhibit similar qualitative behavior: 
complete relaxation for nearest-neighbor impurities and incomplete or strongly delayed relaxation for next-nearest-neighbor impurities. 
The underlying mechanisms leading to this behavior differ between models and include impurity-induced bound states, emergent approximately conserved local observables, and exact cancellation of local and nonlocal dissipation effects. 
\end{abstract} 

\maketitle 

\section{Introduction}
\label{sec:intro}

\begin{figure}[b]
\includegraphics[width=0.7\columnwidth]{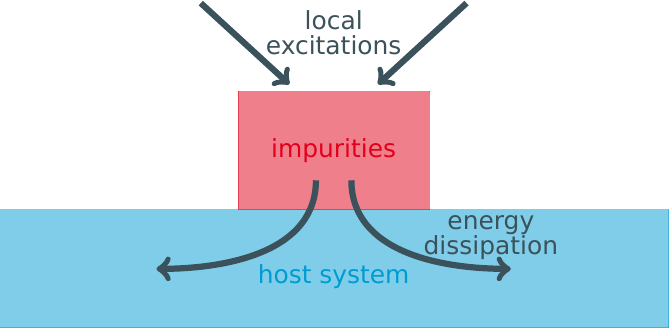}
\caption{
Typical structure of an impurity model.
The impurities are coupled to a much larger host system.
After an initial local excitation the full system, impurities plus host, is expected to relax locally in its ground state in the vicinity of the impurities after the excitation energy is dissipated into the bulk.
}
\label{fig:Introduction_Sketch}
\end{figure}

A small system (``impurity'') in an excited state and coupled to a large bath (``host'') at zero temperature is usually expected to relax over time and to approach its ground state.
After the initial excitation of the impurity, the excess energy is dissipated via the impurity-host coupling and through the coupling of the host degrees of freedom into the bulk of the host system (see Fig.\ \ref{fig:Introduction_Sketch}).
For a system with a macroscopically large number of degrees of freedom subject to the principles of thermodynamics, this dissipation process is irreversible. 
This picture of generic relaxation dynamics explains the interest in exceptional cases, where the system is trapped in a metastable state that does not decay on timescales exceeding by far the typical intrinsic timescales governing the microscopic degrees of freedom. 

Incomplete or delayed relaxation and metastability in impurity-host models \cite{BP02,dVA17} is closely related to incomplete or delayed thermalization of extended lattice models. 
In both cases, much of the interest in metastable states is due to their promise for controlling non-equilibrium dynamics and for related functionalities \cite{dlTKC21}.
Compared to notoriously difficult lattice models, models with single or few, initially excited impurities embedded in a large host represent an interesting class of comparatively simple systems that may hold a key to the understanding of metastability. 
Here, we report on metastable states in the real-time dynamics of three different prototypical system-bath models, an uncorrelated quantum, a classical and a quantum-classical hybrid model. 
In all three cases, the exact dynamics is numerically accessible on long time scales.

In the recent decades, much progress has been made in understanding the thermalization of generic macroscopically large quantum systems, the main paradigm of quantum-statistical physics and the foundation of thermodynamics \cite{Jay57a,Jay57b}. 
Here, an important concept is the eigenstate thermalization hypothesis \cite{Deu91,Sre94,RDO06,SR10,ZS13} as reviewed, e.g., in Ref.\ \cite{Pol11}.
One route to {\em non-thermal} states of quantum-lattice models is via integrability. 
For (one-dimensional) systems with a large number of conserved local observables, the long-time dynamics may result in a state described by a generalized Gibbs ensemble \cite{Jay57,RMO06,RDYO07,CIC12}.
Another route is provided via disorder, either on the single-particle level or via many-body localization \cite{And58,AABS19}. 

This is similar to {\em classical} Hamiltonian dynamics \cite{Arn78}:
It is known that ergodicity and the equivalence between long-time and ensemble averages of observables can be broken in the case of a large number of integrals of motion. 
Violations of ergodicity are found for integrable systems \cite{Arn78} but also for systems parametrically close to integrability \cite{Kol54,Arn63,Mos62,FPUT55} or in systems with glassy dynamics \cite{Jac86,DS01}. 

For quantum-lattice models, too, proximity to an integrable point in parameter space may lead to pre-thermalization and metastability. 
This has been analyzed analytically and demonstrated numerically in several studies \cite{BBW04,MK08,RRBV08,EKW09,KWE11}.

For {\em impurity-host} systems (or open quantum systems) the situation is not very different:
After an initial local excitation of the impurity, one generally expects a relaxation of the reduced density matrix of the impurity to its (canonical) thermal state if the impurity-host coupling is weak \cite{dVA17,vHov54,RRO89,LBS91,GRT00}. 
On the other hand, in the case of band gaps or finite band widths, incomplete relaxation and residual dissipationless dynamics may occur \cite{ZLX+12,XLZFN15}.
Relaxation to non-thermal states may be found in the gapless case for a sufficiently strong impurity-bath coupling 
\cite{XLZFN15,CYS14,ISLN14}.

Recently, a metastable state and incomplete spin relaxation have been observed in a system consisting of a classical impurity spin that is exchange coupled to a spinful Su-Schrieffer-Heeger model at an edge site \cite{EP21}. 
Here, the topological state of the host and the associated presence or absence of a protected edge mode are found to control the relaxation of the classical spin. 
This is a prime example for an impurity system with pre-relaxation dynamics, analogous to pre-thermalization in quantum-lattice models. 
However, the dynamical decoupling of the impurity and the stabilization of the excited state on a long time scale is due to a gapped spectrum for {\em two-particle} excitations, which blocks further energy dissipation.
A similar effect has been observed for a classical spin locally coupled to a one-dimensional half-filled Hubbard model \cite{SRP16a}. 
In this case the Hubbard-$U$ and the narrow spectrum of (quantum) spin excitations control pre-relaxation and metastability. 
Fast but incomplete relaxation to a metastable intermediate excited state, followed by extremely slow complete relaxation is also known from the decay of a local doublon excitation in the Hubbard model at large $U$ \cite{SGJ+10,HP12}, or for a magnetic doublon \cite{RPK19} in the strong-$J$ limit of the Kondo lattice.

Here, we study the exact real-time dynamics by numerical integration of the fundamental equations of motion for three different prototypical impurity models, a quantum, a classical, and a quantum-classical hybrid model. 
All share equivalent geometries, namely a one-dimensional lattice model serving as the bath and two additional impurities which are locally coupled to nearest-neighbor (n.n.) or to next-nearest-neighbor (n.n.n.) sites of the lattice.
Specifically, we study 
(i) a tight-binding quantum model of independent spinless fermions on a lattice with two stub impurities, 
(ii) a classical-spin Heisenberg model with two weakly coupled classical impurity spins, 
and
(iii) a tight-binding quantum model of independent electrons with two classical impurity spins.
The long-time relaxation dynamics initiated by a local excitation of the impurities can by studied numerically for large lattices in all three cases, and in all cases we find qualitatively very similar results:
There is complete relaxation to a time-independent final state in the case of n.n.\ impurities while there is incomplete relaxation or pre-relaxation for n.n.n.\ impurities.
In all three cases the effect can be understood after a thorough theoretical analysis. 
However, it turns out that the uncovered mechanisms are very different.
The different systems are discussed separately in Secs.\ \ref{sec:SIM}, \ref{sec:CHIM}, and \ref{sec:QC}. 
Our conclusions are summarized in Sec.\ \ref{sec:con}.

\section{Stub Impurity Model}
\label{sec:SIM}

\begin{figure}[b]
\includegraphics[width=0.9\columnwidth]{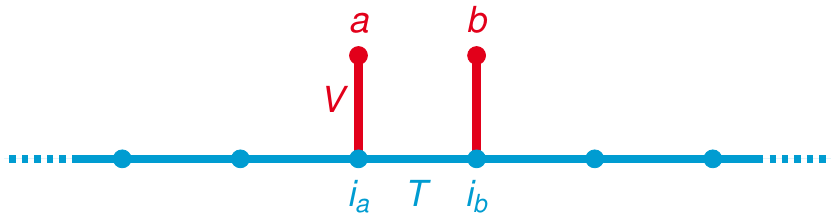}
\caption{
Sketch of the stub impurity model of spinless fermions. 
Two fermionic impurity sites or orbitals are coupled via a hybridization $V$ to a one-dimensional lattice with nearest-neighbor hopping $T$. 
$T=1$ sets the energy scale. 
The host system is at half-filling.
}
\label{fig:Sketch-Stub}
\end{figure}

We start the discussion by considering a tight-binding model with two stub impurities. 
The host system is given by non-interacting spinless fermions on a one-dimensional chain of $L$ sites with open boundaries. 
The nearest-neighbor hopping $T=1$ sets the energy scale. 
The two impurities ($a$ and $b$) are given by two additional sites or orbitals coupling via a hybridization of strength $V$ to the host sites $i_a$ and $i_b$. 
We will consider n.n.\ or n.n.n.\ sites $i_a , i_b$ located at the center of the chain.
A sketch of the system is shown in Fig.\ \ref{fig:Sketch-Stub}. 
The Hamiltonian consists of three terms:
\be
H = H_\text{host} + H_\text{imp} + H_\text{hyb} \; , 
\label{eq:ham}
\ee
the Hamiltonian of the host,
\be
H_\text{host} = -T \sum_{i=1}^{L-1} c_i^\dagger c_{i+1} + \text{H.c.} \; ,
\label{eq:host}
\ee
the impurity sites, 
\be
H_\text{imp} =  \epsilon_f (f^\dagger_a f_{a} + f^\dagger_b f_{b} ) 
\label{eq:HamiltonianStub-imp}
\ee
with on-site energy $\epsilon_f = 0$, 
and the host-impurity hybridization
\ba
H_\text{hyb} = V (c_{i_a}^\dagger f_{a} + c_{i_b}^\dagger f_{b}) + \text{H.c.} \: .
\label{eq:HamiltonianStub-int}
\ea
Here, $c_i$ annihilates a fermion at site $i$, and 
$f_{a}$ annihilates a fermion at impurity site $a$.

We study the real-time dynamics of the system after a quantum quench of the hybridization from zero to a finite value $V$. 
At time $t=0$, the state of the host is assumed to be prepared in its non-degenerate ground state with $N=L/2$ fermions, i.e., half-filling:
\ba
|\Psi_\text{host} (0) \rangle = \prod_k^\text{occ.} c_k^\dagger |\text{vac.}  \rangle \: ,
\ea
where $k$ runs over the one-particle eigenstates of $H_\text{host}$ with one-particle energies $\varepsilon_{k} < 0$. 
Furthermore, the impurity sites $a$ and $b$ are assumed as fully occupied at time $t=0$.
Hence, the initial state of the full system is the state
\ba
|\Psi (t=0) \rangle = f_{a}^{\dagger} f_{b}^{\dagger} |\Psi_\text{host} (0) \rangle \: .
\label{eq:sini}
\ea

As the system is non-interacting, it is sufficient to formulate an equation of motion in terms of the one-particle reduced density matrix $\bm{\rho}$. 
Its elements are defined as 
\ba
\rho_{IJ} (t) = \langle c^\dagger_J c_I \rangle_t \: .
\label{eq:rhoij}
\ea
The indices $I,J$ run over the $L+2$ host and impurity sites: $I,J \in \{1,2,...,L,a,b \}$. 
Initially, at $t=0$, the density matrix $\ff \rho(0)$ has a block-diagonal form with an $L\times L$ block representing the host system and two $1\times 1$-blocks representing the impurities. 
We are interested in the time evolution of the occupation numbers
\ba
n_i(t) \equiv & \langle c^\dagger_i c_i \rangle_{t} = \rho_{ii}(t) \: , \nonumber \\
n_a(t) \equiv  & \langle f^\dagger_a f_a \rangle_{t} = \rho_{aa}(t) \: , \nonumber \\
n_b(t) \equiv  & \langle f^\dagger_b f_b \rangle_{t} = \rho_{bb}(t) \: .
\ea
At $t=0$, we have
\ba
\bm{\rho}_{\text{host}}(t=0) = \Theta(\mu \bm{\mathbb{I}} - \bm{T}_{\rm host} ) \: , 
\ea
where $\Theta$ denotes the Heaviside step function, $\mu=0$ the chemical potential, and $\ff T_{\rm host}$ the hopping matrix of the host system.
Furthermore, $\rho_{aa} (0) = \rho_{bb}(0) =1$.

The time dependence of the density matrix $\ff \rho(t)$ is obtained via the von Neumann equation of motion
\ba
i \frac{d}{dt} \bm{\rho}(t) = \left [\bm{T} , \bm{\rho}(t) \right] \: .
\label{eq:eom_Stub}
\ea
Here, $\bm{T}$ is the hopping matrix of the full system, Eq.\ (\ref{eq:ham}).
The formal solution of Eq.\ (\ref{eq:eom_Stub}) is given by 
\ba
\bm{\rho}(t) = \bm{U} e^{-i\bm{\varepsilon} t} \bm{U}^\dagger \bm{\rho}(0) \bm{U} e^{i\bm{\varepsilon} t} \bm{U}^\dagger \: , 
\label{eq:rho(t)}
\ea
where the diagonal matrix of one-particle eigenenergies $\bm{\varepsilon}$ and the unitary matrix $\ff U$ formed by the one-particle eigenstates of $\ff T$ are obtained by solving the eigenvalue problem
\ba
\ff T \ff U = \ff U \ff \varepsilon \: . 
\label{eq:eigval}
\ea

\begin{figure}[t]
\includegraphics[width=0.9\columnwidth]{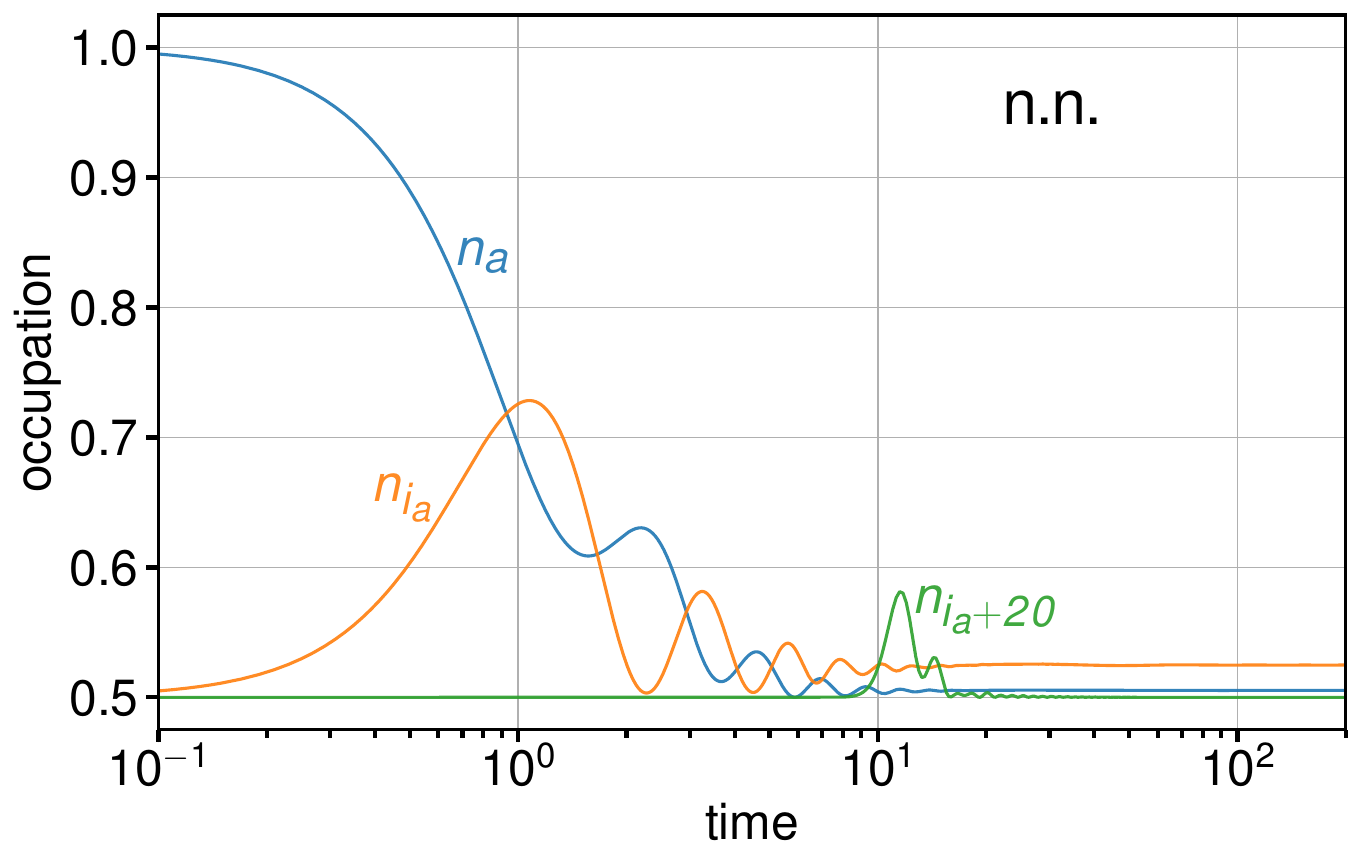}
\caption{
Time dependence of the average occupation number $n_{a}$ of the impurity sites $a$ and of the occupation $n_{i_{a}}$ of the corresponding host site $i_{a}$. 
In addition the occupation $n_{i_{a}+20}$ of a distant site $i_{a}+20$ is displayed.
Calculation for a system with $L=500$ host sites and two stub impurities at the nearest-neighbor positions $i_a=250$ and $i_b=251$. 
Hybridization strength $V=1$. 
Note that inversion symmetry enforces $n_{b}=n_{a}$ and $n_{i_{b}}=n_{i_{a}}$.
The time unit is set by the inverse nearest-neighbor hopping $1/T=1$.
}
\label{fig:stub_nn}
\end{figure}

\begin{figure}[b]
\includegraphics[width=0.99\columnwidth]{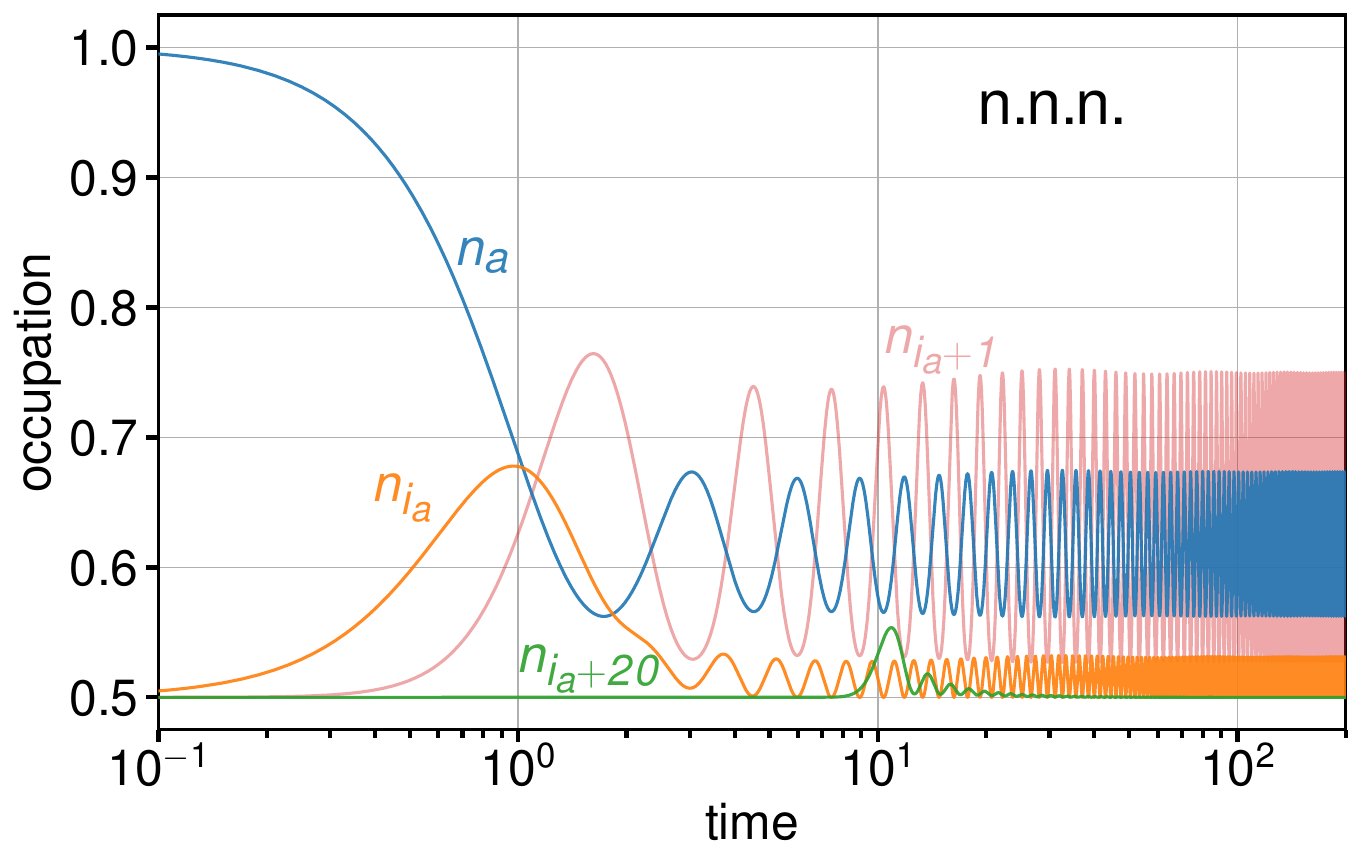}
\caption{
The same as Fig.\ \ref{fig:stub_nn}, but for stub impurities coupled to next-nearest-neighbor (n.n.n.) sites $i_a=249$ and $i_b=251$. 
We also choose $L=499$.
Therewith, inversion symmetry enforces $n_{b}=n_{a}$ and $n_{i_{b}}=n_{i_{a}}$.
}
\label{fig:stub_nnn}
\end{figure}

The $t=0$ state, Eq.\ (\ref{eq:sini}), is not an eigenstate of the full Hamiltonian with $V>0$. 
It instead represents a state that is locally excited, in the vicinity of the impurities.
One naively expects that the local excess energy and the locally enhanced fermion density at the impurity sites are dissipated to the bulk of the system over time and that the system approaches the fully relaxed state that is locally characterized by the ground-state energy density and the ground-state impurity occupations.

Numerical results for a system with $L=500$ host sites and impurities $a$ and $b$ coupling to nearest-neighbor (n.n.) sites $i_a=250$ and $i_b=251$ located symmetrically around the chain center are shown in Fig.\ \ref{fig:stub_nn}.
On a time scale $t \sim 100$, the impurity occupations $n_{a}$ ($=n_{b}$) relax and approach a value $n_{a} \approx 0.505$ close to their ground-state value $n^{\rm (gs)}_{a}=0.5$. 
Similarly the occupations $n_{i_{a}} = n_{i_{b}}$ of the host sites closest to the impurities and also the occupations at more distant sites, e.g., $n_{i_{a}+20}$ relax. 
The characteristic time scale for the dissipation of particles (and of energy) can be read off from Fig.\ \ref{fig:stub_nn} by comparing the dynamics of $n_{i_{a}}$ with that of $n_{i_{a}+20}$.
The total system size ($L=500$) is large enough such that reflections of the propagating wave packets at the open system boundaries do not yet interfere with the dynamics in the vicinity of the impurities -- on the time interval considered. 

The time evolution turns out to be completely different, however, when coupling the two impurities to next-nearest-neighbor (n.n.n) host sites. 
This is demonstrated with Fig.\ \ref{fig:stub_nnn}.
While we do find a fast initial relaxation, the relaxation process stops at $t \sim 2$, and the impurity occupations start to oscillate around a value ($n_{a} \approx 0.625$) that is considerably larger than the ground-state value. 
These oscillations are undamped and persistent (until finite-size effects in form of interference with reflections from the boundaries become important). 

This qualitative difference between n.n.\ and n.n.n.\ impurities, i.e., complete or incomplete relaxation, is likewise observed for impurities at arbitrarily large but odd or even distances, respectively.
Key to this effect is the absence or presence of bound single-particle energy eigenstates of the post-quench Hamiltonian.

Two different types of localized eigenstates can be distinguished:
(i) For each of the two impurities, there is a pair of bound states that split off from the lower and upper edges of the valence band. 
These four high-excitation energy bound states are localized near the impurities with a weight that decays exponentially at large distances. 
The case of strong hybridization $V$ is instructive: 
For $V \to \infty$, the hopping term, Eq.\ (\ref{eq:host}), can be ignored, and the Hamiltonian describes a system of two decoupled dimers with two degenerate eigenstates at $-V$, and two more degenerate states at $+V$ ($V>0$). 
This degeneracy is lifted for finite $T\ll V$, and two bonding-antibonding pairs of bound states are formed, one with negative energies below the bottom of the band and one with positive energies.
As $V$ decreases, the bound states remain localized and centered around the impurities but their weight is increasingly distributed over the lattice.
At $V=1$, only a single state from each of the two pairs remains, a spatially symmetric bound state with negative energy and an antisymmetric bound state with positive energy, while the other two states have merged with the bulk continuum. 
This first type of bound states is generic and thus present for both cases, impurities coupled to n.n.\ and to n.n.n.\ host sites.

(ii) A bound state of a different, second type is present in the case of n.n.n.\ impurities only.
It is given by 
\ba
| \psi_\text{loc} \rangle 
=
\sqrt{\frac{V^2}{V^2 +2T^2}} 
\left [
c_{i_a + 1}^\dagger + \frac{T}{V} \left (f_{a}^\dagger  + f_{b}^\dagger  \right) 
\right ]
| \text{vac.} \rangle 
\: , 
\nonumber \\
\label{eq:locstate}
\ea
where $i_a +1 = i_b -1$ denotes the host site between the sites coupled to the impurities. 
This state has a finite weight on this and on the two impurity sites only, it is ``superlocalized''. 
Furthermore, its eigenenergy, $\varepsilon_{\rm loc}=0$, resides {\em within} the continuum of band states.
This type of states is well known from flat-band systems \cite{Lie89,Tas92,Tas98,Sut86,VMD98,MA17}: 
When coupling a stub impurity to every second host site, the resulting translationally invariant tight-binding lattice model has a unit cell consisting of three sites, and its band structure features a flat band, resulting from superlocalized states, besides two dispersive bands. 

To discuss the impact of bound states on the post-quench relaxation dynamics, we can straightforwardly adapt some concepts developed in Ref.\ \cite{ZS13}.
Accordingly, we decompose the expectation value $O(t) \equiv \langle O(t) \rangle$ of a local operator $O$ with a Heisenberg time dependence as
\ba
O(t) = \overline{O} + \delta O(t)
\: , 
\ea
where the first term is the long-time average, 
\ba
\overline{O} = \lim_{t \to \infty} \frac{1}{t} \int_0^t d\tau \, O(\tau)
\: , 
\label{eq:longav}
\ea
and where the time-dependent second term $\delta O(t)$ is a fluctuation part with a vanishing long-time average.
As a measure for the strength of persistent temporal fluctuations, we consider the long-time average of the absolute square of the fluctuation part:
\ba
\delta_O^2 = \lim_{t\to \infty} \frac{1}{t} \int_0^t d\tau |\delta O(\tau)|^2 \: .
\ea
For $O = O_{IJ} = c^{\dagger}_{J} c_{I}$, see Eq.\ (\ref{eq:rhoij}), the expectation values $O_{IJ}(t) = \rho_{IJ}(t)$ are given by the elements of the one-particle reduced density matrix. 
A straightforward computation yields:
\ba
\delta_{\rho_{IJ}}^2 
= 
\sum_{\mu\nu}^{\mu \neq \nu} |U_{I\mu}|^2 |U_{J\nu}|^2 |\rho_{\mu \nu}(t=0)|^2
\: .
\label{eq:fluctuations}
\ea
Here, $U_{I\mu}$ is the $I$-th component of the $\mu$-th eigen vector of the total hopping matrix, see Eq.\ (\ref{eq:eigval}), and 
\ba
\rho_{\mu \nu}(t=0) = (\bm{U}^\dagger \bm\rho(t=0) \bm{U})_{\mu \nu}
\ea
is an element of the one-particle reduced density matrix in the basis of the eigenstates of the total hopping matrix.
Furthermore, we have assumed that there are no degeneracies and no gap degeneracies in the spectrum of the one-particle eigenenergies.
In our specific case, the spectrum is indeed non-degenerate. 
However, particle-hole symmetry implies the presence of gap degeneracies, i.e., 
$\varepsilon_{\mu} - \varepsilon_{\nu} = \varepsilon_{\mu'} - \varepsilon_{\nu'}$ for $\mu \ne \mu'$ or $\nu \ne \nu'$.
This modifies the derivation that led to Eq.\ (\ref{eq:fluctuations}), see also Ref.\ \cite{ZS13}, and we find
\ba
\delta_{\rho_{IJ}}^2 
& = &
\sum_{\mu\nu}^{\mu \neq \nu} \sum_{\mu'\nu'}^{\mu' \neq \nu'}
\delta_{\varepsilon_{\mu} - \varepsilon_{\nu} , \varepsilon_{\mu'} - \varepsilon_{\nu'}}
U_{I\mu}^{\ast} U_{J\nu} U_{I\mu'} U_{J\nu'}^{\ast}
\nonumber \\
&\times& \rho_{\mu \nu}(t=0) \rho^{\ast}_{\mu' \nu'}(t=0)
\: .
\label{eq:fluctuationscorr}
\ea

The contribution to temporal fluctuations, measured with $\delta_{\rho_{IJ}}^2$ in Eq.\ (\ref{eq:fluctuationscorr}), that stems from {\em extended} eigenstates must vanish for $L\to \infty$, as the components of the eigenvectors are proportional to $L^{-1/2}$ and hence $|U_{i\mu}|^2 |U_{j\nu}|^2 \sim 1 / L^2 \to 0$.
On the contrary, the components $U_{i\mu}$ of a localized eigenstate $\mu$ are independent of $L$ and can be large at some sites $i$, as compared to the components of delocalized eigenstates, such that their contributions to 
$\rho_{\mu \nu}(t=0)$ in Eq.\ \eqref{eq:fluctuationscorr} become significant.

\begin{figure}[t]
\includegraphics[width=0.99\columnwidth]{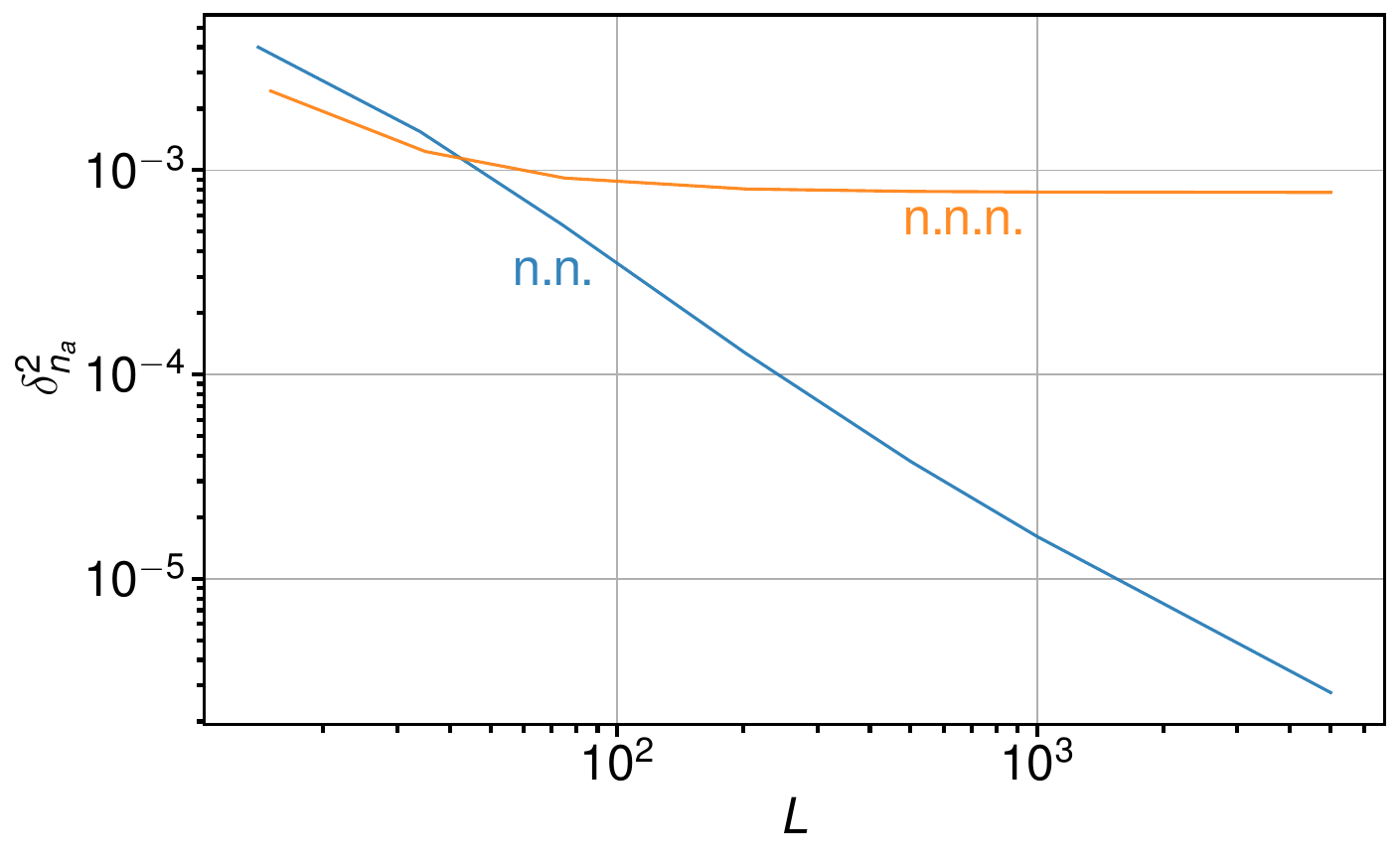}
\caption{
Time-averaged fluctuation of the impurity occupation $\delta_{n_a}^2 = \delta_{\rho_{aa}}^2$, see Eq.\ (\ref{eq:fluctuationscorr}), as a function of the system size $L$ for n.n.\ and n.n.n.\ impurities.
}
\label{fig:fluctuations}
\end{figure}

This is demonstrated with Fig.\ \ref{fig:fluctuations}, where the time-average of fluctuations of the occupation number is shown for one of the impurity sites.
For n.n.\ impurities, $\delta^{2}_{n_{a}}$ decreases with increasing $L$ and eventually vanishes in the thermodynamical limit $L\to \infty$. 
Note that the first type of bound states, discussed above under point (i), does not prevent relaxation due to spatial inversion symmetry, as discussed in the Appendix \ref{sec:nn}.

Contrary, in the case of n.n.n.\ impurities, the strength of the fluctuations is essentially independent of the system size for $L\gtrsim 10^{3}$. 
The non-vanishing temporal fluctuations result from a bound state of the second type, see point (ii) above.
This explains the observed incomplete relaxation for the n.n.n.\ case, see Fig.\ \ref{fig:stub_nnn}. 

In the $L\to \infty$ limit, the effect of gap degeneracies is vanishing in the n.n.\ case. 
In the n.n.n.\ case and for large $L$, the fluctuation $\delta_{n_{a}}^2$, as computed via Eq.\ (\ref{eq:fluctuationscorr}), is smaller by
$3 \cdot 10^{-4}$ (or by about 20\%) compared to the value obtained from Eq.\ (\ref{eq:fluctuations}), i.e., disregarding gap degeneracies.

Note that in the n.n.\ case and for $L=500$, see Fig.\ \ref{fig:stub_nn}, the plateau values for the occupation numbers at $t \approx 100$, i.e., $n_{a} \approx 0.505$ and $n_{i_{a}}\approx 0.525$, are close to but different from their $L\to \infty$ expectation values $n^{\rm(gs)}_{a}=n^{\rm(gs)}_{i_{a}}=0.5$ in the ground state of the post-quench Hamiltonian. 
The latter are fixed by particle-hole symmetry.
For longer propagation times $t_{\rm max} \gg 100$, the plateau in the time evolution (see Fig.\ \ref{fig:stub_nn}) is repeatedly interrupted by revivals (not visible on the time scale in Fig.\ \ref{fig:stub_nn}). 
When computing the long-time averages, Eq.\ (\ref{eq:longav}), including the revivals, we find converged $t\to \infty$ values $\overline{n}_{a} \approx 0.510$ and $\overline{n}_{i_{a}} \approx 0.526$. 
At $L=500$, a propagation time $t < t_{\rm max} \approx 0.5 \times 10^{4}$ has turned out sufficient.

In fact, the dynamics of a system of non-interacting fermions is constrained by the constants of motion $c_{\mu}^{\dagger} c_{\mu}$. 
Hence, the system will relax to a non-thermal state with long-time averages of $n_{a}$ and $n_{i_{a}}$ equal to the averages in the generalized Gibbs ensemble or, equivalently, in the diagonal ensemble (see Ref.\ \cite{ZS13}). 
For an arbitrary operator $O$, the diagonal average is defined as 
\be 
O^{\rm (D)} \equiv \sum_{J} | C_{J} |^{2} O_{JJ}
\: , 
\ee
with $C_{J} = \langle J | \Psi(t=0) \rangle$ and with $O_{JJ} = \langle J | O | J \rangle$.
At $L=500$, for example, the numerical values $n^{\rm (D)}_{a} \approx 0.510$ and $n^{\rm (D)}_{i_{a}} \approx 0.526$ perfectly agree with the above-mentioned long-time averages.
Repeating the computations for larger system sizes (up to $L=5000$) and extrapolating to $L\to \infty$ yields slightly smaller values 
$n^{\rm (D)}_{a} \approx 0.505$ and $n^{\rm (D)}_{i_{a}} \approx 0.525$.

In general, for an integrable system, such as a non-interacting fermion impurity model, the expectation values of local one-particle observables generically do not relax to a thermal state, regardless of the presence or absence of bound states.
However, the presence of bound one-particle eigenstates $\mu$ of the post-quench Hamiltonian is crucial for the question whether there is any relaxation at all or whether the system is trapped in a metastable state.
In the case of n.n.n.\ impurities, there is a superlocalized bound state of the stub impurity model that prevents relaxation and forces the system into a metastable state without any further dissipation.
This explains the qualitatively different relaxation dynamics for n.n.\ and for n.n.n.\ impurities.

\section{Classical Heisenberg Impurity Model}
\label{sec:CHIM}
\begin{figure}[b]
\includegraphics[width=0.8\columnwidth]{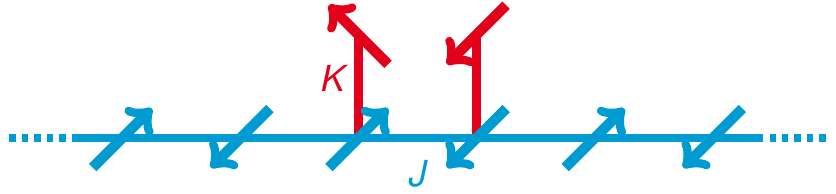}
\caption{
Sketch of a system consisting of two classical impurity spins (red) locally exchange coupled to a one-dimensional classical Heisenberg model (blue spins) with open boundary conditions.
$K$: weak antiferromagnetic local exchange, $J$: antiferromagnetic nearest-neighbor exchange interaction of the host spins. 
}
\label{fig:syssketch}
\end{figure}

A similar effect in the relaxation dynamics is found for a purely classical spin model, i.e., for a one-dimensional Heisenberg model of $L$ classical spins $\ff s_{i}$ ($i=1,...,L$) with nearest-neighbor antiferromagnetic exchange coupling $J>0$, where in addition two classical impurity spins $\ff S_{m}$ ($m=1,2$) are locally exchange coupled to the host spins at sites $i_{1}$ and $i_{2}$.
The geometry is the same as for the stub impurity model, see Fig.\ \ref{fig:syssketch} and compare with Fig.\ \ref{fig:Sketch-Stub}. 
The coupling strength of the local exchange is denoted by $K$. 
We assume that $K\ll J$ and $K>0$, i.e., weak antiferromagnetic exchange interaction. 
The classical Hamilton function is given by
\ba
H = J \sum_{i=1}^{L-1} \bm{s}_i \bm{s}_{i+1} 
+ K \sum_{m=1,2} \bm{S}_m \bm{s}_{i_m}
\: ,
\label{eq:Hamiltonian}
\ea
where the products of vectors are to be understood as dot products. 
The length of the host spins and of the impurity spins is set to $s \equiv |\ff s_{i}| = \frac12$ and $S\equiv |\ff S_{m}| = \frac12$, respectively. 
We consider a lattice with open boundaries. 
The sites $i_{1}$ and $i_{2}$, where the impurity spins are coupled to the host, are assumed to be n.n.\ or as n.n.n.\ sites in the center of the lattice.
The host nearest-neighbor exchange coupling fixes the energy scale, $J=1$, and we assume $K=0.01$ unless otherwise stated.

The equations of motion are easily derived within the classical Hamilton formalism by making use of the spin Poisson bracket \cite{YH80,LD83}. 
They attain the form of Landau-Lifschitz equations \cite{LL35}. 
For the impurity spins we have
\ba
\frac{d}{dt} \bm{S}_m(t) = K \bm{s}_{i_m}(t) \times \bm{S}_m(t)
\: ,
\label{eq:eom1}
\ea
where ``$\times$'' indicates the cross product, 
while for the host spins
\ba
\frac{d}{dt} \bm{s}_i(t) = J \left (\bm{s}_{i-1}(t) + \bm{s}_{i+1}(t) \right ) \times \bm{s}_i(t) \nonumber \\
+ K  \sum_m \left ( \delta_{i i_m} \bm{S}_m(t) \times \bm{s}_i(t) \right ) \: .
\label{eq:eom2}
\ea
It is immediately apparent that the length of each individual spin represents a constant of motion, such that the spin dynamics is constrained to a configuration space given by the $L+2$-fold direct product $\ca S \equiv S^{2} \times \cdots S^{2}$ of 2-spheres with radius $1/2$.
Furthermore, the total energy and, due to the SO(3) spin-rotation symmetry of $H$, the total spin $\sum_m \bm{S}_m + \sum_i \bm{s}_i$ are conserved.
Moreover, the system has an SO(3)-degenerate ground-state manifold, opposed to non-degenerate singlet state of the quantum-spin model \cite{Mar55}.

The spin dynamics is initiated by a parameter quench of the local exchange coupling from zero to a finite value $K$ at time $t=0$. 
We assume that the initial state of the system at $t=0$ is given by one of the ground states for $K=0$. 
For the host spins the ground state is given by an antiferromagnetic N\'{e}el state with respect to an arbitrary axis. 
Specifically, we choose
\ba
  \bm{s}_i (t=0) = (-1)^i s  \bm{e}_z \: .
\ea
The impurity-spin configuration at $t=0$ is taken to be non-collinear, i.e., $\bm{S}_1 = S \bm{e}_x$ and $\bm{S}_2=S\bm{e}_y$. 
This implies that, after switching on $K$ at $t=0$, the spin dynamics is immediately driven by a finite spin torque.

Naively, one again expects that the excitation energy stored in the center of the chain is dissipated into the bulk of the system and that locally, in the vicinity of the impurities, the system approaches a ground-state spin configuration after a sufficiently long propagation time, assuming that the host of the system is sufficiently large to avoid unwanted interactions with excitations back-scattered from the chain boundaries.

\begin{figure}[t]
\includegraphics[width=0.99\columnwidth]{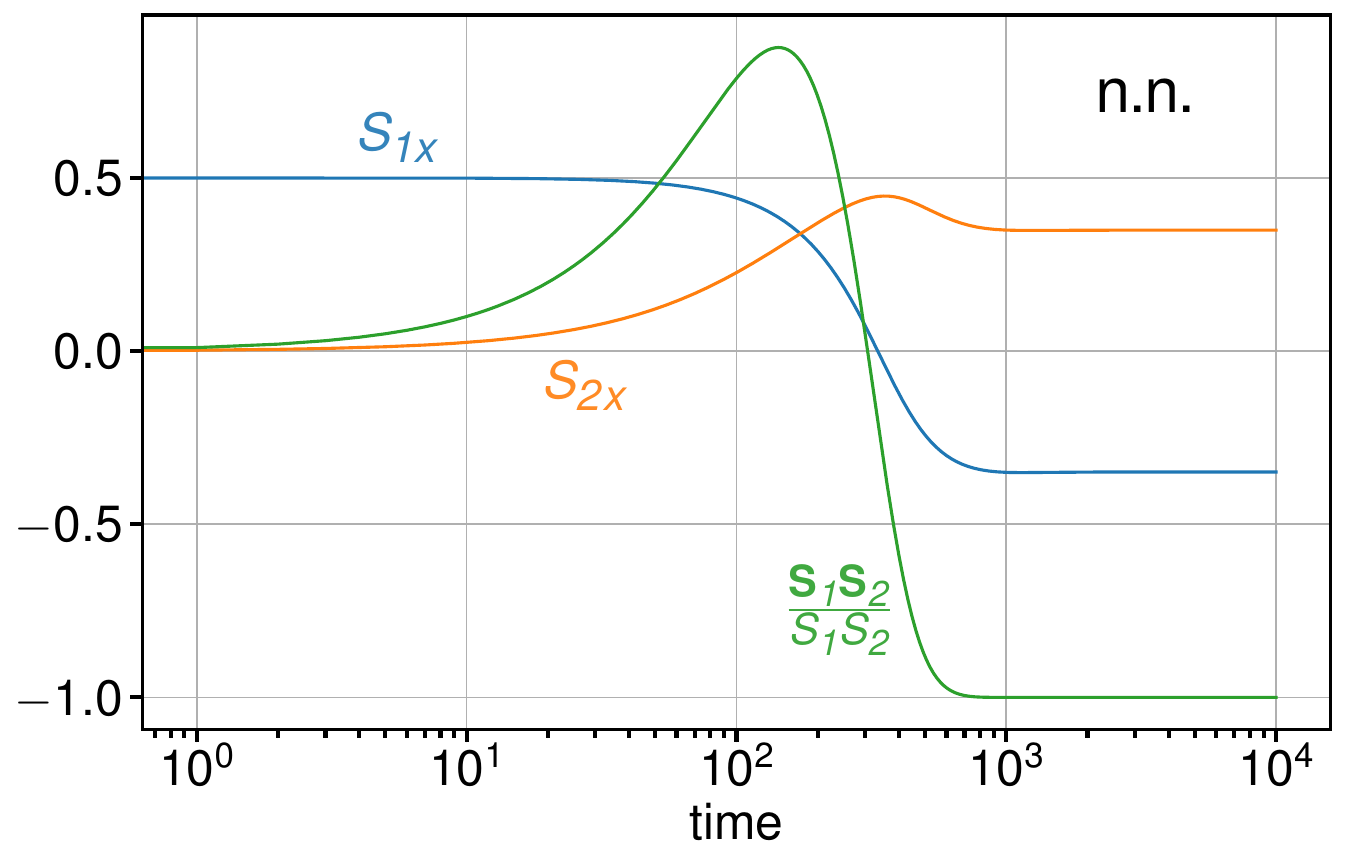}
\caption{
Time evolution of the $x$ components of two classical impurity spins $\ff S_{1}$ and $\ff S_{2}$ and of the cosine of the enclosed angle $\ff S_{1} \ff S_{2}/S_{1}S_{2}$.
Excited state at $t=0$: host spins are in an antiferromagnetic N\'{e}el state aligned to the $z$ axis, 
impurity spins point into the $x$ and $y$ direction, i.e., $\bm{S}_1 = \frac{1}{2} \bm{e}_x$ and $\bm{S}_2 = \frac{1}{2} \bm{e}_y$. 
Geometry parameters: $L=10000$, $i_1=5000$, $i_2=5001$.  
Coupling strengths: $J=1$ and $K=0.01$.
}
\label{fig:relax_nn}
\end{figure}

\begin{figure}[b]
\includegraphics[width=0.99\columnwidth]{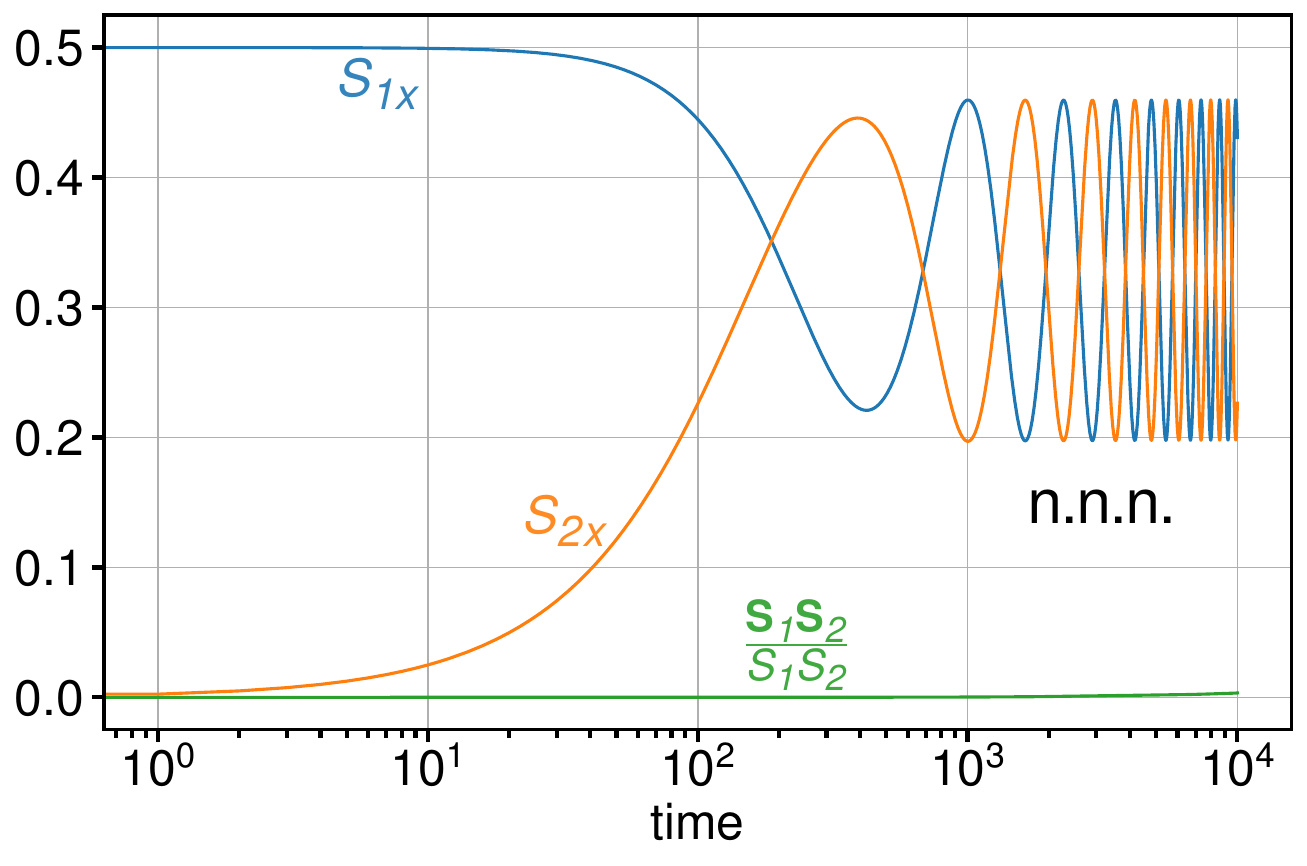}
\caption{
The same as Fig.\ \ref{fig:relax_nn} but for impurity spins coupling to n.n.n.\ host spins.
Geometry parameters: $L=10001$, $i_1=5000$, $i_2=5002$.
}
\label{fig:relax_nnn}
\end{figure}

The equations of motion (\ref{eq:eom1}) and (\ref{eq:eom2}) are easily solved numerically for systems with $L \simeq 10^{4}$ host sites. 
For this system size and for $J=1$, there are no finite-size effects in the form of reflection of spin excitations from the system boundaries up to a propagation time of $t \simeq 10^{4}$.
Fig.\ (\ref{fig:relax_nn}) displays the time dependence of the $x$ components of two impurity spins and of the cosine of the enclosed angle for the case that the impurity spins are locally coupled to nearest-neighbor (n.n.) host spins at the chain center.
We find that after $t \approx 800$ the dynamics has stopped and the system has reached one of its local ground states with an antiferromagnetic impurity-spin configuration and with an antiferromagnetic configuration of the host spins ($\ff S_{1} \uparrow \downarrow \ff S_{2}$) in the vicinity of the chain center.

For two impurity spins coupled to n.n.n.\ sites, however, the time evolution is fundamentally different.
As can be seen in Fig.\ \ref{fig:relax_nnn}, the system does not relax to a local ground state, at least not on the numerically accessible time scale. 
Rather, we find that after a propagation time $t \approx 1000$, the system state becomes trapped in a stationary state, in which the impurity spins precess around a common axis. 
Up to $t=10^{4}$ there is hardly any relaxation to the expected ferromagnetic ($\ff S_{1} \uparrow \uparrow \ff S_{2}$) impurity-spin configuration. 
The angle enclosed by $\ff S_{1}$ and $\ff S_{2}$ starts to deviate only slightly from its initial zero value (see green curve).

We also note that the qualitative difference between the relaxation dynamics for n.n.\ and for n.n.n.\ is not due to the larger distance of the impurity spins in the n.n.n.\ case. 
While the distance between the impurities does have an effect on the relaxation dynamics since it determines the time it takes for the spins to ``communicate'' with each other, this distance dependence turns out as negligible compared to the odd-even effect that is seen in calculations with larger inter-impurity distances $d$. 
In fact, we find quick relaxation for distances $d=1,3,5, ...$ and trapping in a stationary state for $d=2,4,...$, very similar to the results shown in Figs.\ \ref{fig:relax_nn} and \ref{fig:relax_nnn} and analogous to the results for the stub impurity model. 

The observed odd-even effect is actually related to the different ground-state spin configurations, i.e., an antiferromagnetic and a ferromagnetic impurity-spin configuration for n.n.\ and for n.n.n.\ impurities, respectively, and to the small coupling constant $K \ll J$.
This becomes obvious when looking at the time derivative of the scalar product $\ff S_{1} \ff S_{2}$.
From Eq.\ (\ref{eq:eom1}) we get
\ba
\frac{d}{dt} (\bm{S}_1 \bm{S}_2) = K \left ( \bm{S}_1 \times \bm{S}_2 \right ) \left ( \bm{s}_{i_1} - \bm{s}_{i_2} \right ) \: .
\label{eq:S1S2}
\ea
In the following we argue that, for n.n.n.\ impurities, the term on the right-hand side is small, as compared to the inverse of the considered propagation time, in contrast to the n.n.\ case, and that this explains the observed odd-even effect. 
We consider three lines of argument with increasing level of sophistication. 

In the first and crudest approach, we approximate $\bm{s}_{i_2} - \bm{s}_{i_1}$ in an {\em ad hoc} way by its ground-state value. 
For n.n.n.\ impurities, the N\'{e}el ground-state value is $\bm{s}^{(0)}_{i_2} - \bm{s}^{(0)}_{i_1} = 0$, and thus $\ff S_{1} \ff S_{2}$ is a constant of motion that cannot relax to its ground-state value 
$\ff S_{1}^{(0)} \ff S_{2}^{(0)} = 1/4$ (see Fig.\ \ref{fig:relax_nnn}). 
On the other hand, for the n.n.\ case, the right-hand side is generically finite for the N\'{e}el ground state and of the order of $K$. 
In fact, there is a nontrivial dynamics of $\ff S_{1} \ff S_{2}$, and $\ff S_{1} \ff S_{2}$ approaches $\ff S_{1}^{(0)} \ff S_{2}^{(0)} = -1/4$ on a time scale $\Delta t \approx 8 \cdot 10^{2} \sim K^{-1}$ (see Fig.\ \ref{fig:relax_nn}). 

In our second approach we derive an upper bound for $|\bm{s}_{i_1} - \bm{s}_{i_2}|$ for the case of n.n.n.\ impurities. 
For the initial state considered here, where the host spins form a N\'{e}el state, the excitation energy $\Delta E$ is solely stored in the interaction term $\propto K$, see Eq.\ (\ref{eq:Hamiltonian}).
It is given by 
\be
\Delta E 
= 
K \left( \bm{S}_1 \bm{s}^{(0)}_{i_1} + \bm{S}_2 \bm{s}^{(0)}_{i_2} \right ) 
-
K \left( \bm{S}^{(0)}_1 \bm{s}^{(0)}_{i_1} + \bm{S}^{(0)}_2 \bm{s}^{(0)}_{i_2} \right ) 
\: .
\ee
Hence, for an arbitrary initial impurity-spin configuration, the maximum excitation energy is given by $\Delta E_{\rm max} = K$ and is realized for a ferromagnetic alignment of $\bm{S}_m$ and $\bm{s}^{(0)}_{i_m}$.

Assuming that this excitation energy $\Delta E$ is distributed among the bonds between the closest host spins $\bm{s}_{i_1}$, $\bm{s}_{c}$ and $\bm{s}_{i_2}$ only ($c \equiv i_{1}+1=i_{2}-1$), we get 
\ba
\Delta E
&=&
J \left (\bm{s}_{i_1} \bm{s}_{c} + \bm{s}_{c} \bm{s}_{i_2} \right ) - J \left (\bm{s}_{i_1}^{(0)} \bm{s}_{c}^{(0)} + \bm{s}_{c}^{(0)} \bm{s}_{i_2}^{(0)} \right ) 
\nonumber \\
&=&
J (\bm{s}_{i_1} + \bm{s}_{i_{2}}) \bm{s}_{c} 
+ J/2 
\: . 
\label{eq:cspin}
\ea
The right-hand side is at a minimum if the central host spin is $\bm{s}_{c} 
= 
- \frac{1}{2} \frac{\bm{s}_{i_1} + \bm{s}_{i_2}}{ |\bm{s}_{i_1} + \bm{s}_{i_2}|}$. 
With this we find
\ba
\Delta E 
\ge
\frac{J}{2} - \frac{J}{2}  |\bm{s}_{i_1} + \bm{s}_{i_2}|
\: . 
\ea
Using $\ff s_{1}^{2}=\ff s_{2}^{2}=\nicefrac14$ and the parallelogram law, $(\ff s_{1} + \ff s_{2})^{2} + (\ff s_{1} - \ff s_{2})^{2} = 1$, we get
\ba
  |\bm{s}_{i_1} - \bm{s}_{i_2}|^{2}
  \le
  4 \frac{\Delta E}{J} - 4\frac{\Delta E^{2}}{J^{2}} 
  \: . 
\ea
With the above argument, $\Delta E \le \Delta E_{\rm max} = K$, we find
\ba
|\bm{s}_{i_1} - \bm{s}_{i_2}| \leq 2 \sqrt{\frac{K}{J}}
\: .
\label{eq:CHIM-estimation}
\ea
This upper bound is a very conservative estimate as in the course of time the excitation energy will be further dissipated to the bulk of the system, and thus $|\bm{s}_{i_1} - \bm{s}_{i_2}|$ will be even smaller. 
We conclude that for $K\ll J$, the small available excitation energy of order $K$ very much restricts the host spin dynamics. 
Via Eqs.\ (\ref{eq:S1S2}) and (\ref{eq:CHIM-estimation}) this imply that $\ff S_{1} \ff S_{2}$ is almost conserved if $\ff S_{1}$ and $\ff S_{2}$ couple to n.n.n.\ host spins.

Our third approach is based on a linearization of the equations of motion. 
We start from Eqs.\ \eqref{eq:eom1} and \eqref{eq:eom2} and substitute 
$\bm{S}_m = \bm{S}_{m}^{(0)} + \delta \bm{S}_m$ and $\bm{s}_i = \bm{s}_i^{(0)} + \delta \bm{s}_i$, 
where $\bm{S}_{m}^{(0)}$ and $\bm{s}_i^{(0)}$ are ground-state spin orientations while $\delta \bm{S}_m$ and $\delta \bm{s}_i$ denote the deviations from the ground state. 

\begin{widetext}

Linearization of Eqs.\ \eqref{eq:eom1} and \eqref{eq:eom2} yields: 
\be
\dot{\bm{S}}_m = 
J \left[ 
\frac{K}{J} \bm{s}_{i_m}^{(0)} \times \bm{S}_m + \frac{K}{J} \bm{s}_{i_m} \times \bm{S}^{(0)}_m 
+
\mathcal{O} \left (\frac{K^2}{J^2} \right )
\right]
\label{eq:leom1}
\ee
and
\be
\dot{\bm{s}}_i = J \Big[ \left( \bm{s}_{i-1}^{(0)} + \bm{s}_{i+1}^{(0)} \right) \times \bm{s}_i
+
  \left( \bm{s}_{i-1} + \bm{s}_{i+1} \right) \times \bm{s}_i^{(0)} 
+   \sum\limits_{m=1}^{2} \delta_{i i_m} \left (\frac{K}{J} \bm{S}_m^{(0)} \times \bm{s}_i + \frac{K}{J} \bm{S}_m \times \bm{s}_i^{(0)} \right) 
+ \mathcal{O} \left( \frac{K^2}{J^2} \right) \Big]
\: ,
\label{eq:leom2}
\ee
where we used 
\ba
|\delta \ff S_m| 
= 
\mathcal{O} \left ( \frac{\Delta E}{K} \right)
 = \mathcal{O} \left ( 1 \right) 
\: , 
\qquad
|\delta \ff s_i| 
=  
\mathcal{O} \left ( \frac{\Delta E}{J} \right) 
= \mathcal{O} \left ( \frac{K}{J} \right) 
\label{eq:dsOrder}
\ea
to estimate the magnitude of the neglected terms. 
We see that linearization of the equations of motion is possible although $|\delta \ff S_m| = \mathcal{O} \left ( 1 \right)$ is not necessarily small. 
The reasoning is the same that led to Eq.\ \eqref{eq:CHIM-estimation}, i.e., the maximum values for $|\delta \ff S_m|$ and $|\delta \ff s_i|$ are limited by the available initial excitation energy $\Delta E \le K$.
Impurity spins contribute on the order of $K$ to the total energy, while host spins contribute on the order of $J$. 
The estimates (\ref{eq:dsOrder}) are well supported by our numerical results underlying Fig.\ \ref{fig:relax_nnn}.

We proceed by computing the time derivative of $\bm{S}_1 \bm{S}_2$ within the linearized theory.
With Eqs.\ (\ref{eq:leom1}), (\ref{eq:leom2}) we find
\ba
 \frac{d}{dt} (\bm{S}_1 \bm{S}_2) 
&=& J \left [\frac{K}{J} \left (\bm{S}_1 \times \bm{S}_2 \right )  \left ( \bm{s}_{i_1}^{(0)} - \bm{s}_{i_2}^{(0)} \right ) + \frac{K}{J} \left ( \bm{S}_2 \times \bm{s}_{i_1} \right )  \bm{S}_1^{(0)} + \frac{K}{J} \left ( \bm{S}_1 \times \bm{s}_{i_2} \right )  \bm{S}_2^{(0)} \right ] + J \mathcal{O} \left (\frac{K^2}{J^2} \right ) 
\: .
\label{eq:conq2}
\ea
For n.n.n.\ impurity spins, the first term on the right-hand side vanishes since $\bm{s}^{(0)}_{i_1} = \bm{s}^{(0)}_{i_2}$. 
Furthermore, we can write $\bm{S}^{(0)} = \bm{S}_1^{(0)} = \bm{S}_2^{(0)}$. 
Therewith we find
\ba
\frac{d}{dt} (\bm{S}_1 \bm{S}_2 )
=  J\bm{S}^{(0)} \left (\frac{K}{J} \bm{S}_2 \times \bm{s}_{i_1} + \frac{K}{J} \bm{S}_1 \times \bm{s}_{i_2} \right ) + J \mathcal{O} \left (\frac{K^2}{J^2} \right ) \text{.}
\label{eq:conq3}
\ea
With $\bm{S}_i = \bm{S}^{(0)} + \delta \bm{ S}_i$ and $\bm{s}_m = \bm{s}_m^{(0)} + \delta \bm{ s}_m$,  exploiting that ground-state spin configurations are collinear, and finally using Eqs.\ (\ref{eq:dsOrder}), 
one has
\ba
\frac{d}{dt} (\bm{S}_1 \bm{S}_2 )
=
J \bm{S}^{(0)} \left [ \frac{K}{J}
\delta \bm{ S}_2 \times \delta \bm{ s}_{i_1} + \frac{K}{J} \delta \bm{ S}_1 \times \delta \bm{ s}_{i_2} + \mathcal{O} \left (\frac{K^2}{J^2} \right )
\right ]
= J \mathcal{O} \left (\frac{K^2}{J^2} \right )
\: .
\label{eq:conq5}
\ea
This means that, {\em within the linearized theory}, $(d/dt)( \bm{S}_1 \bm{S}_2)$ must be considered as zero and that $\ff S_{1} \ff S_{2}$ is a constant of motion with a correction of the same order of magnitude as the linearization error only.

\end{widetext}

In the case of n.n.\ impurity spins, restarting from Eq.\ (\ref{eq:conq2}) with a completely analogous calculation   but with an antiferromagnetic ground-state alignment $\bm{s}^{(0)}_{i_1} = - \bm{s}^{(0)}_{i_2}$, one finds
\be
\frac{d}{dt} (\bm{S}_1 \bm{S}_2 )
=
J \bm{S}^{(0)} \left [ 
2  \frac{K}{J}  \delta \bm{ S}_1 \times \delta \bm{S}_{2}
\right ]
+ 
J \mathcal{O} \left ( \frac{K^2}{J^2} \right )
\: ,
\label{eq:conq6}
\ee
i.e., there is a nontrivial dynamics on an energy scale that is by an order of magnitude larger than the linearization error, such that, {\em even within the linearized theory}, $\ff S_{1} \ff S_{2}$ cannot be considered as a constant of motion.

We have also studied the dynamics beyond the weak-coupling regime. 
For $K$ and $J$ of the same order of magnitude, one finds a relaxation of $\ff S_{1} \ff S_{2}$ already after a very short propagation time of $t \simeq 100$ for both, the case of n.n.\ and of n.n.n.\ impurity spins.

For the weak-coupling regime $K\ll J$, we conclude that after an initial local excitation of n.n.n.\ impurity spins, these show an anomalous relaxation dynamics. 
There is almost no relaxation of $\ff S_{1} \ff S_{2}$, i.e., the enclosed angle is almost a constant of motion, on a time scale of about $t\sim 10^{4}$. 
This must be contrasted with the case of n.n. impurity spins, where complete relaxation is reached after a propagation time of $t \simeq 800$. 
In contrast to the stub-impurities model discussed above, there is no local symmetry of the (classical) Hamiltonian that would lead to a conserved local observable. 
(Quasi-)conservation of $\ff S_{1} \ff S_{2}$ is rather emerging in the course of time. 
After a certain pre-relaxation process ($t \simeq 10^{3}$) with a sufficient dissipation of energy and spin, the system state has evolved sufficiently close to one of the ground states {\em locally}, i.e., in the vicinity of the impurities, such that the further dynamics is very well captured by linearized equations of motion.
Indeed, within the linearlized theory, $\ff S_{1} \ff S_{2}$ is strictly conserved. 
Its validity range, however, is not only controlled by the a weak local exchange $K \ll J$ but also by the propagation time. 
Residual perturbative deviations from the linear dynamics accumulate over time, such that complete relaxation of the system, also for the n.n.n.\ case, is expected on a long time scale. 
In fact, indications for full long-time relaxation are seen at $t \simeq 10^{4}$ in Fig.\ \ref{fig:relax_nnn}. 

\section{Quantum-classical impurity model}
\label{sec:QC}

In the case of the quantum-classical impurity model, we again find a qualitatively very similar effect in the relaxation dynamics. 
However, to explain the observed incomplete relaxation, it turns out again that a different methodological approach is necessary. 

\begin{figure}[t]
\includegraphics[width=0.99\columnwidth]{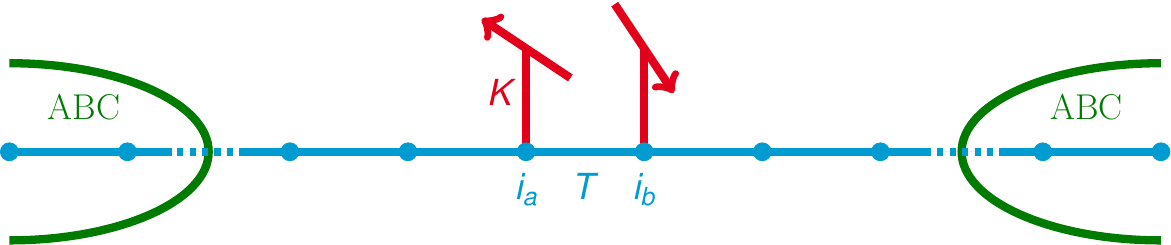}
\caption{
Sketch of the quantum-classical hybrid model: 
Two classical impurity spins (red) are locally exchange coupled to a system of conduction electrons on a one-dimensional lattice (blue). 
$K$: antiferromagnetic local exchange, 
$T$: nearest-neighbor hopping. 
Green: absorbing boundary conditions (ABC).
}
\label{fig:SC_SysSketch}
\end{figure}

We consider a spinful single-orbital tight-binding model on a one-dimensional lattice of $L$ sites with hopping between nearest neighbors $T$, where in addition two classical impurity spins $\ff S_{m}$ ($m=1,2$) are locally exchange coupled to the local electron spins $\ff s_{i} = \frac{1}{2}\sum_{\sigma \sigma'} c_{i \sigma}^\dagger \bm{\tau}_{\sigma \sigma'} c_{i \sigma'}$ at sites $i=i_{m}$ of the lattice via an antiferromagnetic exchange interaction $K$. 
Here, $\bm{\tau}$ denotes the vector of Pauli matrices, and $\sigma=\uparrow, \downarrow$ is the electron  spin projection.
A sketch of the system is shown in Fig.\ \ref{fig:SC_SysSketch}.
The geometry is the same as for the previous models. 
As for the classical Heisenberg model, we assume that $K$ is weak and can be treated perturbatively. 
The quantum-classical Hamiltonian is 
\ba
H = -T \sum_{ij}^\text{n.n.} \sum_{\sigma} c^\dagger_{i\sigma} c_{j\sigma} + K \sum_m \bm{S}_m \bm{s}_{i_m}
\: .
\ea
We choose $T=1$ to fix the energy (and time) scale and an antiferromagnetic exchange $K>0$.
As for the stub impurity model, we consider half-filling, i.e., $N=L$ electrons. 

The equations of motion \cite{Elz12,SP15} couple the classical and the quantum sector of the theory. 
For the classical impurity spins, we obtain Landau-Lifshitz-type equations,
\ba
\frac{d}{dt} \bm{S}_m &=& K \langle \bm{s}_{i_m} \rangle_t \times \bm{S}_m(t)
\: ,
\label{eq:eoms}
\ea
similar to Eq.\ (\ref{eq:eom1}). 
Here, $\langle \bm{s}_{i_m} \rangle_t$ is the expectation value in the $N$-electron state $| \Psi(t) \rangle$ at time $t$. 
For a given classical spin configuration at time $t$, the quantum system is uncorrelated, and hence its real-time dynamics is completely described by the one-particle reduced density matrix $\ff \rho(t)$ with elements
\ba
\rho_{i\sigma i' \sigma'} (t) = \langle \Psi(t)| c_{i'\sigma'}^\dagger c_{i\sigma} | \Psi(t) \rangle 
\: . 
\ea
Its equation of motion is essentially the same as for the stub-impurity model, see Eq.\ (\ref{eq:eom_Stub}), but the hopping matrix $\ff T$ replaced by the time-dependent effective hopping matrix $\ff T^{\rm (eff)}(t)$, which includes the classical impurity spins as time-dependent external parameters:
\be
T^{\rm (eff)}_{i\sigma i' \sigma'}(t) 
=
\delta_{\sigma\sigma'} T_{ii'} 
+
\frac{K}{2} \delta_{ii'} \sum_{m=1,2} \delta_{ii_{m}} \ff \tau_{\sigma \sigma'} \ff S_m(t)
\: .
\label{eq:teff}
\ee

We study the time evolution of the full system starting at $t=0$ from an initial state where the two impurity spins are in an excited non-collinear configuration (as in the classical Heisenberg impurities case, $\bm{S}_1 = S \bm{e}_x$ and $\bm{S}_2=S\bm{e}_y$ with $S=\frac12$), while the electron system is in its ground state corresponding to this spin configuration. 
The excitation energy stored in the vicinity of the impurities is dissipated to the bulk of the electron system on a time scale that, even for $K$ of the order of $T$, typically exceeds by far the time scale that is numerically accessible when using open boundaries and when reflections of propagating excitations from the boundaries of the system shall be avoided. 
Since the propagation is essentially ballistic, unphysical reflections from the boundaries that disturb the dynamics near the impurities will occur at time $t \sim L / T$, i.e., one would have to work with effective hopping matrices of very high matrix dimension.
For this reason and as indicated in Fig.\ \ref{fig:SC_SysSketch}, we employ so-called absorbing boundary conditions (ABC), which have been developed and extensively tested previously, see Ref.\ \onlinecite{EP20}.
Apart from the conserving von Neumann term, the resulting equations of motion contain a dissipative term and are given by
\be
i \frac{d}{dt} \bm{\rho}(t) 
=
\left [ \bm{T}_\text{eff}(t),\bm{\rho}(t) \right ]
- i \{\bm{\gamma} , \bm{\rho}(t) - \bm{\rho}_0 \}
\: .
\label{eq:eome}
\ee
Here, $\ff \rho_0$ is the initial ground-state one-particle reduced density matrix,
$\{ \cdot, \cdot \}$ denotes the anticommutator, 
and $\ff \gamma$ is a diagonal matrix controlling the dissipation rate. 
It has nonzero entries only for the outermost two ``absorbing'' sites on both sides of the chain.
For the concrete computations, we have fixed the values for $\gamma_{1}=\gamma_{L}$ and $\gamma_{2}=\gamma_{L-1}$, as in Ref.\ \cite{EP20}, by comparing the resulting spin dynamics using ABC with the exact spin dynamics, i.e., without the dissipative term in Eq.\ (\ref{eq:eome}). 
This has been done for a shorter propagation time of $t = 5 \cdot 10^{2}$ and a larger system size such that reflections from the boundaries are avoided.
We find perfect agreement with $\gamma_{1}=0.230$ and $\gamma_{2}=0.115$. 
However, the results are quite insensitive to the precise choice. 
As compared to the system studied in Ref.\ \cite{EP20}, the optimal parameters are smaller, because the spin dynamics is much slower.

\begin{figure}[t]
\includegraphics[width=0.99\columnwidth]{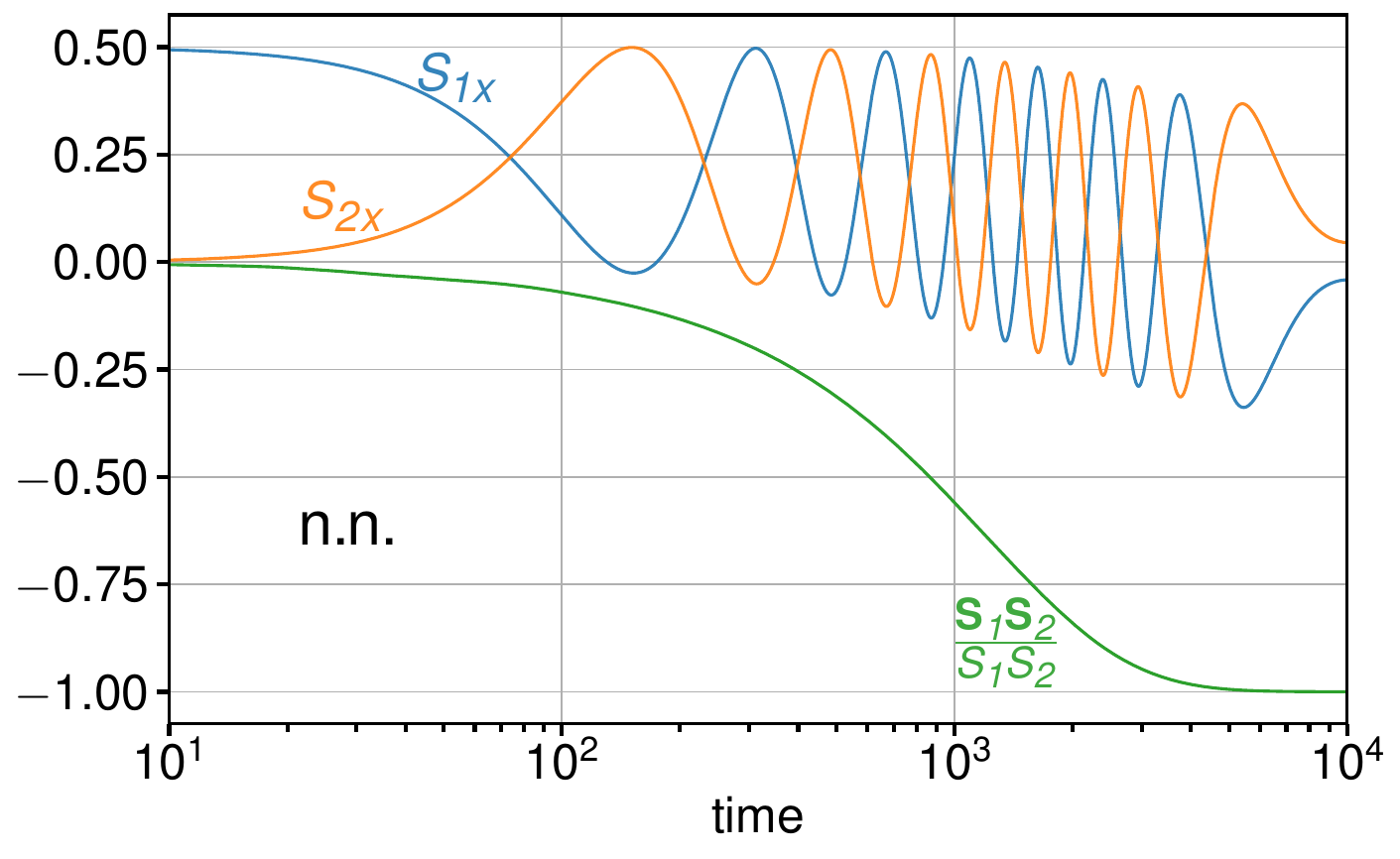}
\caption{
Time evolution of the $x$ components of the two classical impurity spins $\ff S_{1}$ and $\ff S_{2}$ (blue and orange) exchange coupled to the conduction-electron system at neighboring sites $i_{1}=34$ and $i_{2}=35$ at the center of a chain with $L=68$ sites. 
Green: cosine of the angle enclosed by $\ff S_{1}$ and $\ff S_{2}$.
Initial excited state at time $t=0$: host electron system in its ground state corresponding to the initial (excited) impurity-spin configuration $\bm{S}_1 = \frac{1}{2} \bm{e}_x$ and $\bm{S}_2 = \frac{1}{2} \bm{e}_y$. 
Further parameters: $T=1$, $K=1$. 
Absorbing boundary conditions with nonzero diagonal elements $\gamma_{2}=\gamma_{L-1}=0.115$ and $\gamma_{1}=\gamma_{L}=0.230$ (see text).
}
\label{fig:SC_nn}
\end{figure}

The impurity-spin dynamics for nearest-neighbor spins as obtained by solving the coupled equations of motion (\ref{eq:eoms}) and (\ref{eq:eome}) is displayed in Fig.\ \ref{fig:SC_nn}.
At short times $t\lesssim 10^{3}$ there is a pronounced precession dynamics with a small frequency $\omega \sim 0.01$. 
This is explained by the indirect RKKY exchange \cite{RK54,Kas56,Yos57} mediated by the electron system, which is rather weak, 
even for an exchange interaction of $K=T=1$.
On a longer time scale $t\sim 10^{4}$, the system shows complete relaxation, and the two spins reach their antiferromagnetic ground-state configuration (see green line in the figure).

\begin{figure}[t]
\includegraphics[width=0.99\columnwidth]{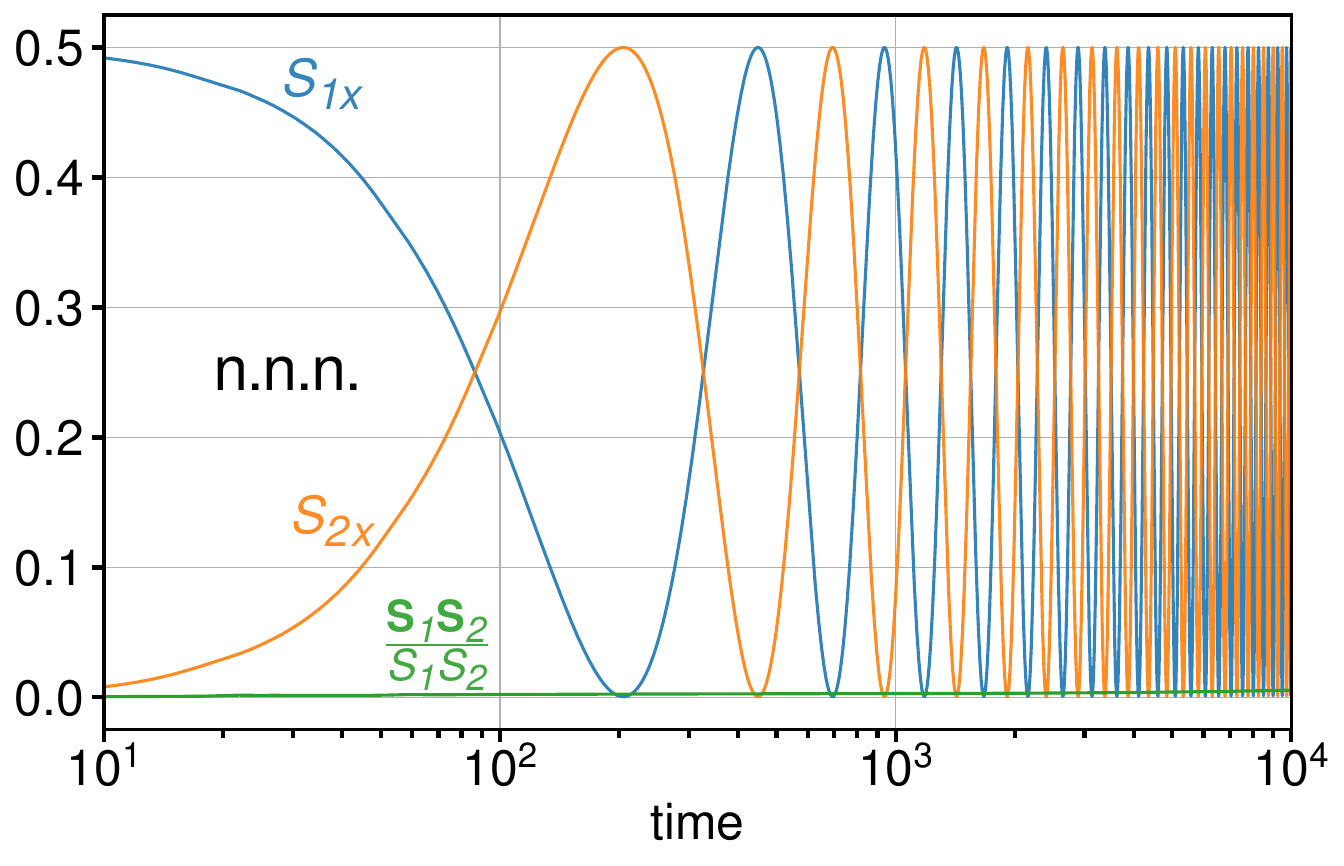}
\caption{
The same as in Fig.\ \ref{fig:SC_nn} but for next-nearest-neighbor impurity spins at $i_1=34$, $i_2=36$ of a chain with $L=69$ sites in total.
}
\label{fig:SC_nnn}
\end{figure}

For impurity spins coupled to n.n.n.\ sites, see Fig.\ \ref{fig:SC_nnn}, the same precessional motion is found, but with an even smaller precession frequency. 
This reflects the smaller RKKY exchange due to the increased distance between the spins. 
However, the real-time dynamics is qualitatively different, as there is hardly any relaxation to the ferromagnetic ground-state spin configuration visible on the numerically accessible time scale.
At time $t=10^{4}$, the angle enclosed by $\ff S_{1}$ and $\ff S_{2}$ deviates by less than 1\% from its initial value only.
We conclude that, as for the other models studied, the system is trapped in an intermediate stationary state and that complete relaxation, if at all, takes place on a still much longer time scale.

For smaller $K$ (not shown here), the time evolution is essentially the same in qualitative terms. 
The only notable difference is that the dynamics is even slower, i.e. characterized by smaller precession frequencies and longer relaxation times.

None of the explanations for incomplete relaxation used for the previously discussed systems is easily applicable to the quantum-classical model.
The linearization of the coupled equations of motion is not helpful to identify possible local conserved observables, and in fact is not informative due to the large number of $\sim 4L^{2}$ of degrees of freedom in the quantum sector, i.e., the density-matrix elements.

With linear-response theory \cite{ON06,BNF12,UMS12,SP15,EP21}, we choose a different approach. 
Conceptually, this is limited to the weak exchange-coupling regime and directly addresses the dynamics of the classical impurity spins. 
For weak $K$, the expectation value of the local spin at site $i_{m}$ can be obtained via the Kubo formula as 
\be
  \langle \ff s_{i_{m}} \rangle_{t} = K \sum_{m'=1,2} \int_{0}^{t} dt' \, \chi_{mm'}(t') \ff S_{m'}(t-t') 
  \: .
\label{eq:kubo}
\ee
Here, the response function, the $K=0$ retarded magnetic susceptibility of the electron system is isotropic
$\chi_{m\alpha ,m'\alpha'}(t) = \delta_{\alpha\alpha'} \chi_{mm'}(t)$ and independent of the spatial direction 
$\alpha = x,y,z$. 
It is given by
\ba
\chi_{m\alpha ,m'\alpha'}(t) = -i \Theta(t) e^{-\eta t} 
\langle [ s_{i_{m} \alpha}(t), s_{i_{m'} \alpha'}(0) ] \rangle
\: , 
\ea
where $\langle \cdots \rangle$ is the $K=0$ ground-state expectation value and where $\eta$ is a positive infinitesimal.
A straightforward computation yields
\ba
\chi_{mm'}(t) 
&=&
\Theta(t) e^{-\eta t} \mbox{Im} 
\Big[
\left( e^{-i \ff T t} \Theta(\ff T - \mu) \right )_{i_{m} i_{m'}} 
\nonumber \\
&\times &
\left( e^{i \ff T t} \Theta(\mu-\ff T) \right)_{i_{m'}i_{m}} 
\Big]
\: .
\label{eq:chi}
\ea
Note that the chemical potential $\mu = 0$ at half-filling.
For the evaluation of Eq.\ (\ref{eq:chi}), we consider large systems with up to $L = 10^{5}$ sites and choose periodic boundary conditions. 
Hence, the hopping matrix is diagonalized as $\ff T = \ff U \ff \varepsilon \ff U^{\dagger}$, where the unitary matrix $\ff U$ with elements $U_{ik} = e^{ik R_{i}}/\sqrt{L}$ describes Fourier transformation from lattice sites $i$ to wave ``vectors'' $k$ in the first Brillouin zone. 
The entries of the diagonal matrix are given by the tight-binding dispersion $\varepsilon_{k} = - 2T \cos(k)$. 
We define $A_{i_m i_{m'}}(k,k') = U^\dagger_{k i_m} U_{i_mk'} U^\dagger_{k' i_{m'}} U_{i_{m'} k}=L^{-2}e^{i(k' - k)(i_m-i_{m'})}$. 
Furthermore, we write $\Delta \varepsilon_{kk'} = \varepsilon_k - \varepsilon_{k'} = - 2T (\cos(k)- \cos(k'))$ for short.
This yields:
\ba
\chi_{mm'}(t) 
&=& 
- \frac{i}{2} \Theta(t) e^{-\eta t}  
\sum_k^{\rm occ.} \sum_{k'}^{\rm unocc.}  
A_{i_{m}i_{m'}}(k,k') 
\nonumber \\
& \times & \left( e^{i\Delta \varepsilon_{kk'}t} - e^{-i\Delta \varepsilon_{kk'}t} \right)
\: , 
\ea
or, after Fourier transformation from time to frequency space, the frequency-dependent susceptibility 
\ba
\chi_{mm'}(\omega) 
&=& 
\frac{1}{2} 
\sum_k^{\rm occ.} \sum_{k'}^{\rm unocc.} A_{i_{m}i_{m'}}(k,k') 
\nonumber \\ 
& \times &
\left(
\frac{1}{\omega - \varepsilon_{k} + \varepsilon_{k'} + i \eta} 
- 
\frac{1}{\omega + \varepsilon_{k} -  \varepsilon_{k'} + i \eta} 
\right)
\: . 
\nonumber \\ 
\label{eq:chiom}
\ea
Note that we have the symmetry $\chi_{mm'}(\omega) = \chi_{m'm}(\omega)$ for the nonlocal elements $m\ne m'$, while the local elements are $m$ independent, $\chi_{mm}(\omega) = \chi_{m'm'}(\omega)$, due to translation invariance.
The representation (\ref{eq:chiom}) is well suited to compute the Gilbert damping:
\be
\alpha_{mm'} 
=
-i K^2 \frac{\partial}{\partial \omega} \chi_{mm'}(\omega) \rvert_{\omega=0} 
\label{eq:alpha}
\ee
and the RKKY indirect magnetic exchange
\be
J_{mm'} = K^2 \chi_{mm'} (\omega = 0)
\: ,
\label{eq:rkky}
\ee
which determine the effective equations of motion for the classical spin dynamics
(see Ref.\ \onlinecite{SP15}):
\be
\dot{\bm{S}}_m = \sum_{m'} J_{mm'} \bm{S}_{m'} \times \bm{S}_m 
+ 
\sum_{m'} \alpha_{mm'} \bm{S}_m \times \dot{\bm{S}}_{m'} 
\: .
\label{eq:llg}
\ee

\begin{figure}[t]
\includegraphics[width=0.99\columnwidth]{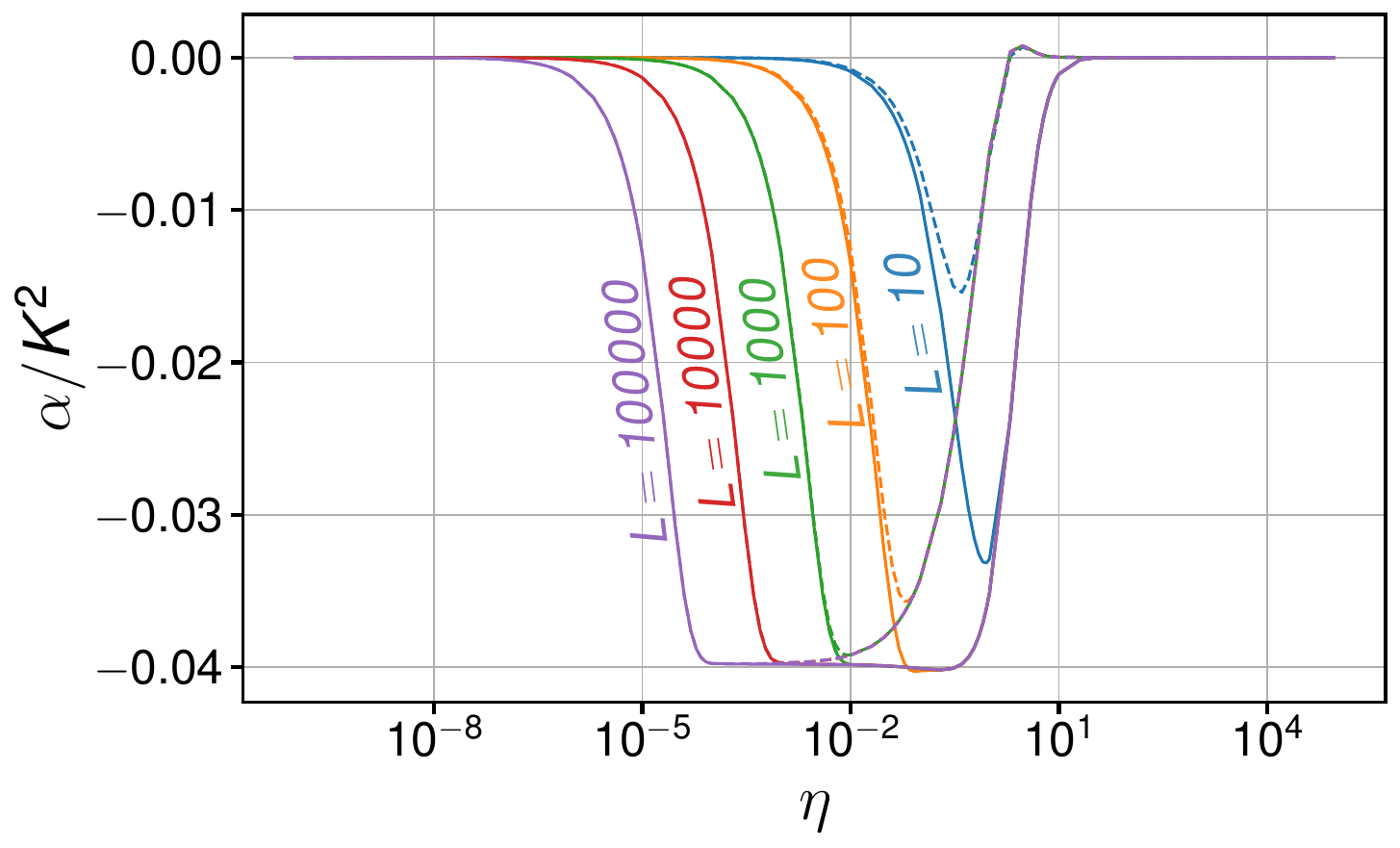}
\caption{
Local ($m=m'$, solid lines) and nonlocal next-nearest-neighbor Gilbert damping ($m\ne m'$, dashed lines) $\alpha/K^{2}$ as function of $\eta$ for different system sizes $L$ as indicated. 
Results for large systems with periodic boundary conditions, $T=1$. 
}
\label{fig:alpha-eta-L}
\end{figure}

In practice, the results of various calculations for different system sizes $L$ as well as for different $\eta$ must be extrapolated to obtain physical results in the thermodynamic limit $L\to \infty$ and in the limit $\eta \to 0$. 
Here, it is important to take the thermodynamic limit first. 
This is demonstrated with Fig.\ \ref{fig:alpha-eta-L}, where the local, $\alpha_{mm}$ and the nonlocal (n.n.n.) Gilbert damping $\alpha_{mm'}$ ($m\ne m'$) are shown as function of $\eta$ for different $L$. 
We start the discussion with the local damping (solid lines).
First, we see that for any {\em fixed} $\eta \gtrsim 10^{-4}$, the values for the local Gilbert damping nicely converge with increasing $L$. 
System sizes of about $L=100,000$ are sufficient for numerical convergence unless even smaller values of $\eta$ are considered.
Second, the converged values $\lim_{L\to \infty} \alpha_{mm}$ become independent of $\eta$ with decreasing $\eta$, once $\eta$ is sufficiently small. 
We find a rather precise value $\lim_{\eta\to 0}\lim_{L\to \infty} \alpha_{mm} \approx -0.0398 \, K^{2}$. 
Here, we note that taking the limits in the opposite order yields the unphysical result
$\lim_{L\to \infty} \lim_{\eta\to 0} \alpha_{mm} = 0$.
This is easily understood: 
For any finite $L$, the spectrum of one-particle energies is gapped. 
Close to $\omega=0$, the finite-size gap is $\delta \approx 2\pi / L$. 
This implies that for $\eta \lesssim \delta \approx 2 \pi / L$, the Gilbert damping must start to deviate from its physical value and approach $\alpha_{mm}=0$, as there is no damping in a finite system.
For the practical calculations, it has turned out that when fixing the ``infinitesimal'' at $\eta \approx 2 \pi /L$, it is sufficient to control the convergence with respect to $L$ only.

Our considerations for computing the Gilbert damping likewise apply to the case of n.n.n.\ spins. 
There is, however, an important physical result that can be read off from Fig.\ \ref{fig:alpha-eta-L}: 
In the case of n.n.n.\ spins, the converged value for the nonlocal Gilbert damping (see dashed lines) is exactly the same as the local damping, i.e., 
$\lim_{\eta\to 0}\lim_{L\to \infty} \alpha_{mm'} \approx -0.0398 \, K^{2}$ for both, $m=m'$ and $m \ne m'$ within numerical accuracy. 
We note that a similar result for the nonlocal Gilbert damping has been found for metallic ferromagnets with quadratic energy-momentum dispersion \cite{RON24}.

The equality between the local and the nonlocal damping has in fact important consequences for the spin dynamics, as can be easily seen when rewriting Eq.\ (\ref{eq:llg}) explicitly for two classical spins but with a single Gilbert damping constant $\alpha \equiv \alpha_{11}=\alpha_{12}=\alpha_{12}=\alpha_{21}$:
\ba
\dot{\bm{S}}_1 &=& J \bm{S}_{2} \times \bm{S}_1 
+ \alpha \bm{S}_1 \times \dot{\bm{S}}_{1} 
+ \alpha \bm{S}_1 \times \dot{\bm{S}}_{2} 
\: , \nonumber \\
\dot{\bm{S}}_2 &=& J \bm{S}_{1} \times \bm{S}_2 
+ \alpha \bm{S}_2 \times \dot{\bm{S}}_{2} 
+ \alpha \bm{S}_2 \times \dot{\bm{S}}_{1} 
\: .
\label{eq:llg1}
\ea
Note that only the nonlocal RKKY exchange coupling $J \equiv J_{mm'}=J_{m'm}$ ($m\ne m'$) enters the equations.
We immediately see that the total impurity spin $\ff S_{\rm tot} = \ff S_{1} + \ff S_{2}$ and thus $\ff S_{1} \ff S_{2}$ are constants of motion, 
as in the case of the classical Heisenberg impurity model, see Sec.\ \ref{sec:CHIM}.
This implies that there is no relaxation to the ground-state spin configuration at all. 

So far we have discussed the case of n.n.n.\ impurity spins, where as a consequence of $\alpha_{mm} = \alpha_{mm'}$ ($m\ne m'$) there is no relaxation to the ground-state spin configuration.
While for n.n.\ impurity spins the local Gilbert damping $\alpha_{mm} \approx - 0.398 K^2$ stays the same, we find, on the other hand, $\alpha_{mm'} \approx  0.0021K^2$ ($m\ne m'$) for the nonlocal Gilbert damping. 
The signs are such that a solution of Eq.\ (\ref{eq:llg}) must approach the ground state, i.e., in the n.n.\ case an antiferromagnetic spin configuration.
This is consistent with the computed positive RKKY exchange coupling $J_{12}\approx0.0342 K^2$
($H_{\rm RKKY} = J_{12} \ff S_{1} \ff S_{2}$) for the n.n.\ case.
On the contrary $J_{12} \approx -0.0189K^2$ for the n.n.n.\ case with ferromagnetic ground-state spin configuration.

Returning to the n.n.n.\ case, the equality of the local and the nonlocal damping can be understood analytically. 
Using Eqs.\ (\ref{eq:chiom}) and (\ref{eq:alpha}) one finds
\ba
\alpha_{mm'}
&=& 
\frac{i}{2} K^2 
\sum_k^{\rm occ.} \sum_{k'}^{\rm unocc.} A_{i_{m}i_{m'}}(k,k') 
\nonumber \\ 
& \times &
\left(
\frac{1}{(- \varepsilon_{k} + \varepsilon_{k'} + i \eta)^{2}} 
- 
\frac{1}{(\varepsilon_{k} -  \varepsilon_{k'} + i \eta)^{2}} 
\right)
\: . 
\nonumber \\ 
\label{eq:al}
\ea
Nonzero contributions to the double sum are obtained from wave vectors close to the Fermi points, i.e., for $k,k' = \pm \pi/2 + \ca O(1/L)$ only (note that $\varepsilon(k=\pm \pi/2) = 0 = \mu$).
Contrary, for $k , k' = \pm \pi/2 + \ca O(1)$, the imaginary infinitesimal can be disregarded, since we may take $\eta = \ca O(1/L)$, as argued above, and thus $\ca O(1/L) = \eta \ll |\varepsilon_{k} -  \varepsilon_{k'} | = \ca O(1)$, and the two fractions in Eq.\ (\ref{eq:al}) cancel exactly in the thermodynamical limit.

It is thus sufficient to analyze the contributions from $k=\pm \pi/2 + \delta k$ and $k'=\pm \pi/2 + \delta k'$ with $\delta k, \delta k' = \ca O(1/L)$ for $L\to \infty$ and show that these give the same result for $i_{m'}=i_{m}$ and for $i_{m'}=i_{m}+2$.
The $i_{m}, i_{m'}$ dependence of the Gilbert damping is due to the weight factor $A_{i_{m}i_{m'}}$ only.
We therefore focus on $A_{i_{m}i_{m'}}$.
Its imaginary part does not contribute to the double sum in Eq.\ (\ref{eq:al}). 
For the discussion of the real part, we first consider $k=\pi/2 + \delta k$ and $k'=\pi/2 + \delta k'$:
\be
\mbox{Re} \, A_{i_{m}i_{m'}}(k,k')  = \frac{1}{L^2} \cos \left((-\delta k + \delta k')(i_m - i_{m'}) \right)
\: .
\ee
Now, if $i_{m'}=i_{m}$, we have $\mbox{Re} \, A_{i_{m}i_{m'}}(k,k') = L^{-2}$, and if $i_{m'}=i_{m}+2$ we get
\be
\mbox{Re} \, A_{i_{m}i_{m'}}(k,k') 
=
\frac{1}{L^2} \left(1+\ca O \left (\frac{1}{L^2} \right)\right)
\: , 
\label{eq:Aii}
\ee
and, hence, $A_{i_{m}i_{m'}}(k,k') = A_{i_{m}i_{m}}(k,k') + \ca O(L^{-2})$. 
Analogously, this also holds for $k=\pi/2 + \delta k$ and $k'=-\pi/2 + \delta k'$ and for $k=-\pi/2 + \delta k$ and $k'=\pi/2 + \delta k'$ and $k=-\pi/2 + \delta k$ and $k'=- \pi/2 + \delta k'$. 
This concludes our argument. 

\begin{figure}[t]
\includegraphics[width=0.99\columnwidth]{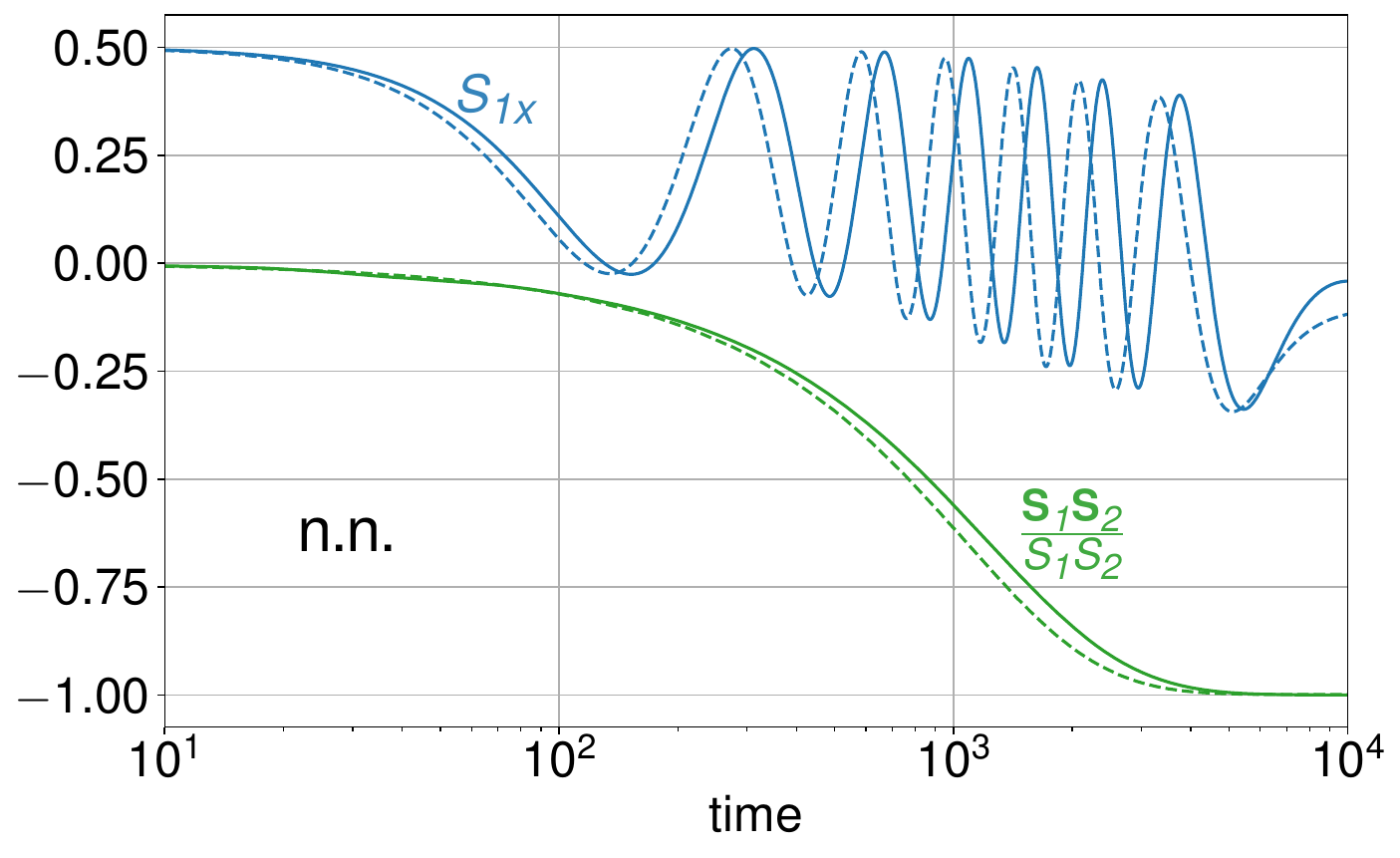}
\caption{
Comparison between the full spin dynamics (solid lines), as obtained from the exact equations of motion (\ref{eq:eoms}) and (\ref{eq:eome}) and absorbing boundary conditions, and linear-response spin dynamics (dashed lines), as obtained from Eq.\ (\ref{eq:llg}) with numerically determined parameters 
$\alpha_{11} = \alpha_{22} = -0.0398$, $\alpha_{12}=\alpha_{21} = 0.0021$, and $J_{12} = J_{21} = 0.0342$, see Eqs.\ (\ref{eq:alpha}) and (\ref{eq:rkky}), respectively. 
Time evolution of the $x$ component of $\ff S_{1}$ (blue) and cosine of the angle enclosed by $\ff S_{1}$ and $\ff S_{2}$ (green) for the case of n.n.\ impurity spins.
All other parameters as in Fig.\ \ref{fig:SC_nn}. 
In particular, $K=T=1$.
}
\label{fig:gnn}
\end{figure}

\begin{figure}[t]
\includegraphics[width=0.99\columnwidth]{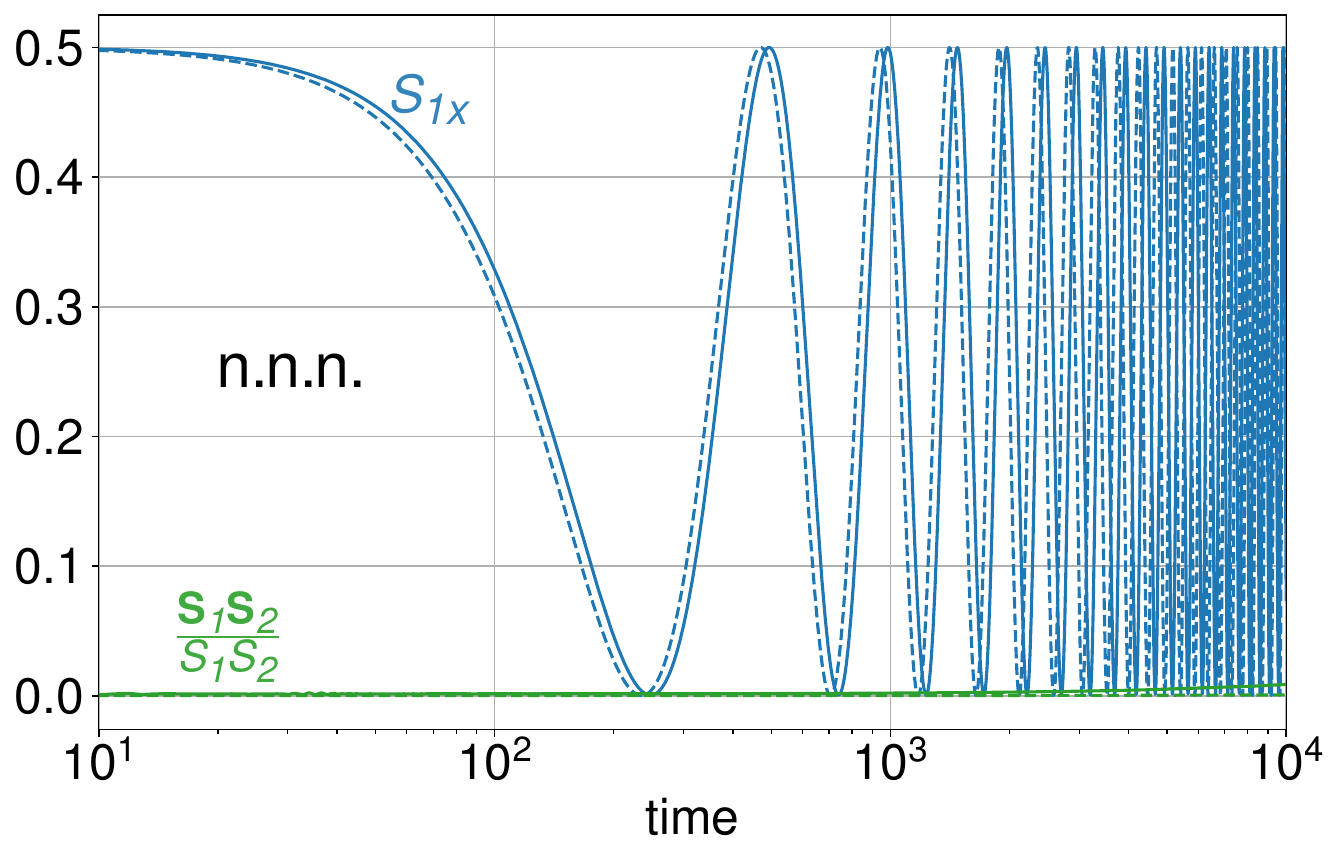}
\caption{
Comparison between the full spin dynamics and linear-response spin dynamics
(calculated damping and exchange parameters: $\alpha_{11} = \alpha_{22} = \alpha_{12} = \alpha_{21} = -0.0398$, and $J_{12} = J_{21} = -0.0189$) as in Fig.\ \ref{fig:gnn} but for next-nearest-neighbor impurity spins.
All other parameters as in Fig.\ \ref{fig:SC_nnn}. 
In particular, $K=T=1$.
}
\label{fig:gnnn}
\end{figure}

The argument extends to arbitrary $i_{m}$, $i_{m'}$ if $i_{m'} - i_{m}$ is even, but fails at macroscopic distances $i_{m'} - i_{m} = \ca O(L)$. 
It is also invalid for n.n.\ impurities and, more generally, for odd distances between the impurities, because
for $k=\pi/2 + \delta k$ and $k'=-\pi/2 + \delta k'$, e.g., we have $A_{i_{m}i_{m'}}(k,k')=1$ for $i_{m}=i_{m'}$ and
$A_{i_{m}i_{m'}}(k,k') = -1 + \ca O(L^{-2})$ for n.n.\ $i_{m}$, $i_{m'}$, and for odd distances.

Since our explanation of the incomplete relaxation is based on perturbative-in-$K$ linear-response theory, it is necessary to compare corresponding results with those of the full theory (using absorbing boundary conditions), Eqs.\ (\ref{eq:eoms}) and (\ref{eq:eome}).
We choose $K=T$ for this comparison. 
This provides us with a comparatively fast spin dynamics. 
Results are displayed in Figs.\ \ref{fig:gnn} and \ref{fig:gnnn} for the cases of n.n.\ and n.n.n.\ impurity spins. 

We find a slight phase offset in the precessional motion for the n.n.n.\ case (Fig.\ \ref{fig:gnnn}). 
On the logarithmic time scale, this offset is constant.
Furthermore, at late times $t \sim 10^{4}$, a tiny deviation of the angle enclosed by the two spins from its initial $t=0$ value is visible in the results from the full theory, hinting towards complete relaxation on a much longer time scale. 
This is missing in the linear-response approach. 

For the n.n.\ case, where the spin dynamics is much more complicated, the perturbative method also does an almost perfect job, see Fig.\ \ref{fig:gnn}. 
While we observe the same but slightly larger phase shift and a slightly longer relaxation time, all the qualitative features of the spin dynamics are fully captured. 

We conclude that linear-response approach itself, i.e., perturbation theory in $K$ is quite reliable even for comparatively strong $K\sim T$ and errors accumulating up to a time scale $t \sim 10^{4} / T$ do not affect the qualitative trend of the spin dynamics. 
This also holds for the typical additional approximations that are necessary to arrive at Eqs.\ (\ref{eq:alpha}) and (\ref{eq:rkky}), i.e., weak retardation effects and time independence of the Gilbert damping, see Refs.\ \cite{BNF12,SP15}.
All in all the numerical results demonstrate that the proposed mechanism based on the analysis of the nonlocal Gilbert-damping term in fact captures the essence of the incomplete relaxation.

\section{Conclusions}
\label{sec:con}

Using numerical simulations, we have studied the exact real-time dynamics of three different prototypical one-dimensional model systems with two impurities coupled locally to nearest-neighbor or to next-nearest-neighbor sites of the host. 
In all cases we considered an initial state with a local excitation at or near the impurities.
The unifying theme of all three models studied is the conservation (or the approximate conservation) of observables that are localized in the vicinity of the impurities.
Furthermore, in all cases, the presence of these (quasi) conserved observables depends on the system geometry, and in all cases this is crucial for the relaxation dynamics.

The independent-electron tight-binding quantum model with two stub impurities is conceptually the simplest.
Due to the lack of interactions, it is integrable; its real-time dynamics is strongly constrained by a macroscopically large number of conserved observables.
This implies that local one-particle observables do not relax to their ground-state values but to a non-thermal GGE-like state respecting the constraints.
However, it depends crucially on the geometry, i.e., on the relative position of the impurities, whether or not a complete relaxation to a time-independent state for $t \to \infty$, for an infinitely extended system, actually occurs.
In the case of impurities coupled to n.n.n.\ sites, we find persistent oscillations up to the numerically accessible time scale, i.e., before unwanted reflections from the boundaries set in.

The classical Heisenberg model with two locally exchange coupled classical impurity spins shows very similar behavior. 
For n.n.\ impurity positions, there is a fast and complete relaxation to the ground state impurity-spin configuration. 
In contrast, in the n.n.n.\ case, an (almost) undamped oscillatory spin dynamics is found, when the local exchange coupling $K$ is weak compared to the host exchange $J$. 
Complete relaxation to the ground state configuration is not observed on the numerically accessible time scale. 
However, the numerical data indicate that complete relaxation is possible on a much longer time scale, so that the system actually exhibits pre-relaxation.
This is an essential difference from the non-interacting quantum system. 

Qualitatively the same results are found for the quantum-classical hybrid model with two classical impurity spins locally exchange coupled to an independent-electron system on the one-dimensional lattice, i.e., fast complete relaxation to the ground-state spin configuration in the n.n.\ case, while in the n.n.n.\ case and after a fast pre-relaxation, a metastable intermediate state is formed, in which the impurity spins undergo an 
(almost) undamped oscillation.
This intermediate state is stable up to $t \gtrsim 10^{4}$ in units of the inverse hopping. 
Again, we assume that the impurity-host coupling, the local exchange $K$, is sufficiently weak.

An explanation for the observed very different behavior for n.n.\ vs. n.n.n.\ geometries, common to all three models, does not seem obvious. 
In fact, quite different theoretical concepts have been put forward as explanations:

The incomplete relaxation of the quantum system with n.n.n.\ impurities is due to the presence of a superlocalized energy eigenstate bound to the impurities and thus due to a local observable commuting with the Hamiltonian. 
The superlocalized state is reminiscent of the states forming flat bands in tight-binding models on lattices with characteristic geometries.

The metastability of the classical spin model, on the other hand, could be traced back to an {\em approximately} conserved local observable, reminiscent of explanations for the pre-thermalization of interacting lattice models parametrically close to an integrable point.
In fact, we had to assume that $K \ll J$, which places the model parametrically close to the trivial $K=0$ point.
Here, the weak-coupling limit has allowed us to linearize the equations of motion and thus to understand the approximate conservation law.
This is remarkable because the fluctuations $\delta \bm{S}_m$ of the impurity spins around their ground-state configuration $\bm{S}_{m}^{(0)}$ are not at all small, since the metastable state is far from the ground state. 
Rather, the argument can be based on the fact that the excitation energy is of the order of $K \ll J$, and that each impurity spin contributes of the order of $K$ to the total energy, as opposed to the host spins which contribute of the order of $J$.
We note that this reasoning seems possible only for impurity models.

The analysis of the real-time dynamics of the quantum-classical hybrid model is much more complicated, since a simple linearization of the equations of motions is not very feasible and not justified.
However, the limit of weak local exchange coupling $K \ll T$ could be exploited in another way, namely by time-dependent perturbation or linear-response theory.
This turns out to be reliable even up to intermediate couplings $K \sim T$ and propagation times $t \lesssim 10^{4}$ in units of the inverse hopping parameter.
{\em Within} the linear-response framework, the {\em stability} of persistent oscillations in the spin dynamics in the case of n.n.n.\ impurities is nicely explained by a perfect cancellation of local with nonlocal Gilbert damping constants.
However, the exact dynamics obtained numerically clearly indicates that, beyond the perturbatively accessible time scale, the non-equilibrium steady state is actually metastable and that there is further relaxation on a much longer time scale.

While it seems to make no qualitative difference for the relaxation dynamics if quantum degrees of freedom are replaced by classical ones or vice versa, the geometry is a crucial factor.
Common to all three models studied is the bipartite system geometry, i.e., the one-dimensionality of the host lattice with nearest-neighbor couplings between host sites, and the local host-impurity coupling. 
Admittedly, we expect that next-nearest-neighbor host or nonlocal host-impurity couplings (hopping or spin-exchange couplings) break the (meta)stability of the non-equilibrium state in the case of n.n.n.\ impurities and lead to (faster) complete relaxation.
However, parametric proximity to the bipartite geometry, e.g., weak nonlocal host-impurity couplings, should still lead to a significantly different relaxation dynamics between impurities at n.n.\ and n.n.n.\ positions.
Thus, we believe that our results provide valuable insights for our understanding of metastable states and the control of non-equilibrium dynamics.
The study of the relaxation dynamics of impurities coupled to a host system on a two- or three-dimensional lattice is one of the promising avenues of further research.

Furthermore, it would be very interesting to study the relaxation dynamics for systems including quantum rather than classical spins or, more generally, for correlated quantum impurity models and to check the robustness of the results against quantum fluctuations. 
It is quite conceivable that also in such systems the geometry plays a crucial role for the existence of (approximately) conserved local quantities.
Of course, it will be technically more challenging to reach the relevant time scales.
For one-dimensional systems, however, matrix-product-state approaches \cite{Sch11} seem to be suitable to study relaxation dynamics, see e.g. Ref.\ \cite{LRP22}.

\acknowledgments
This work was supported by the Deutsche Forschungsgemeinschaft (DFG) through the Cluster of Excellence ``Advanced Imaging of Matter'' - EXC 2056 - project ID 390715994, and through the Sonderforschungsbereich 925 (project B5), project ID 170620586, and through the research unit QUAST, FOR 5249 (project P8), project ID 449872909.

\appendix

\section{Complete relaxation in the case of nearest-neighbor stub impurities}
\label{sec:nn}

We consider the model Eq.\ (\ref{eq:ham}) with n.n.\ stub impurities for $V=1$ (corresponding to Figs.\ \ref{fig:stub_nn} and \ref{fig:fluctuations}), where there are two bound states at energies outside the band continuum.
Let us refer to these bound eigenstates of the post-quench Hamiltonian as $\mu=b_{1}$ and $\mu= b_{2}$, respectively.
When calculating the fluctuations via Eq.\ \eqref{eq:fluctuations}, the only contributions to the double sum that are non-vanishing in the thermodynamical limit $L\to \infty$ are due to these bound eigenstates. 
Hence, there are essentially only two terms to be taken into account: $\mu=b_1, \nu=b_2$ and $\mu = b_2, \nu = b_1$. 
Consider a corresponding element of the initial one-particle reduced density matrix at time $t=0$ in the basis of the eigenstates of the post-quench Hamiltonian:
\ba
\rho_{b_1 b_2}(t=0) = \sum_{IJ} U_{b_1 I}^\dagger \rho_{IJ}(t=0) U_{J b_2} \: .
\ea
To exploit the mirror symmetry of the system under reflection at the chain center, we define 
$\tilde{I} = L - I$, if $I$ is a host site, while $\tilde{I}$ shall refer to the other impurity site, if $I$ is an impurity site.
With this notation, we can symmetrize the summation as follows
\ba
\rho_{b_1 b_2}(0)
= \frac{1}{2} \sum_{IJ} \left [ U_{b_1 I}^\dagger \rho_{IJ}(0) U_{J b_2} + U_{b_1 \tilde{I}}^\dagger \rho_{\tilde{I}\tilde{J}} (0)U_{\tilde{J} b_2} \right ] \: .
\nonumber \\
\ea
The node theorem in quantum mechanics requires that the lowest-energy state, say $\bm{U}_{b_1}$, be symmetric under reflection, i.e., $U_{b_1 I}=U_{b_1 \tilde{I}}$, while the highest-energy state $\bm{U}_{b_2}$ is antisymmetric, $U_{\tilde{J} b_2} = - U_{J b_2}$. 
This immediately implies $\rho_{b_1 b_2}(0) = 0$. 
We conclude that there are no flucturations surviving the thermodynamical limit $L\to \infty$ in the case n.n.\ impurities.
Note that this argument is invalid for the n.n.n.\ case.
The reason is that the inversion symmetry is different due to a different inversion center, i.e., there is an invariant site $i_{a}+1=i_{b}-1$, opposed to the n.n.\ case.

\end{document}